\documentclass[journal]{IEEEtran}

\usepackage[utf8]{inputenc} % For typing special characters 	
\usepackage[T1]{fontenc}

\usepackage{amstext}
\usepackage{amsfonts}
\usepackage{amsmath,amssymb,amsthm}
\usepackage{color}
\usepackage{framed}

\usepackage{graphicx}
\usepackage[font=footnotesize]{caption}
\usepackage{subcaption}
\usepackage{epstopdf}
\usepackage{bibentry}
\usepackage{cite}
\usepackage{comment}
\usepackage[T1]{fontenc}
\usepackage{mathtools}
\usepackage{algpseudocode}
\usepackage{algorithm}
\newcommand{\ieb}{\begin{IEEEeqnarray}{rCl}}
\newcommand{\ien}{\end{IEEEeqnarray}}
\usepackage{array} 
\newcolumntype{L}{>{$}l<{$}} % math-mode version of "l" column type

\DeclareMathOperator{\Tr}{Tr}
\DeclareMathOperator*{\argmin}{arg\,min}
\DeclareMathOperator*{\argmax}{arg\,max}

\title{Hybrid Precoding in Cooperative Millimeter Wave Networks}
\author{Chao Fang ,
		Behrooz Makki , \IEEEmembership{Senior Member,~IEEE,
		} Jingya Li  and
		Tommy Svensson , \IEEEmembership{Senior Member,~IEEE,
	}
\thanks{This work was supported in part by VINNOVA (Swedish Government Agency for Innovation Systems) within the VINN Excellence Center ChaseOn.
	
	C. Fang and T. Svensson are with the Dept. of Electrical Engineering, Chalmers University of Technology, Gothenburg, Sweden (e-mail: \{fchao, tommy.svensson\}@chalmers.se).}
\thanks{B. Makki and J. Li are with Ericsson Research, Ericsson AB, Gothenburg, Sweden (e-mail:\{behrooz.makki, jingya.li\}@ericsson.com ).}
}
\begin{document}
	\maketitle
	\begin{abstract}
		
	%	Millimeter wave (MmWave) communications with hybrid precoding
	%	are able to achieve gigabits-per-second data rate. The hybrid precoding architectures, which use fewer radio frequency chains than the number of antennas, 
	%	achieve low cost and low complexity for mmWave systems with large antenna arrays. However, because mmWave signals are sensitive to blockages, most hybrid precoding techniques that focus on the performance of single serving base station (BS) may not satisfy users' quality-of-service requirements.

        In this paper, we study the performance of cooperative millimeter wave (mmWave) networks with hybrid precoding architectures. 
        Considering joint transmissions and BS silence strategy, we propose hybrid precoding algorithms which minimize the sum power consumption of the base stations (BSs), for both fully- and partially-connected hybrid precoding (FHP and PHP, respectively) schemes, for single-carrier and orthogonal frequency-division multiplexing systems. We reformulate the analog precoding part as an equal-gain transmission problem, which only depends on the channel information, and the digital precoding part as a relaxed convex semidefinite program subject to per-user quality-of-service constraints that gives the optimal sum power consumption in terms of the BS silence strategy. In order to reduce the complexity of the hybrid precoding algorithm with optimal BS silence strategy, we propose a sub-optimal hybrid precoding algorithm that iteratively put BSs with small power into the silent mode.
		The simulation results
        show that, depending on the parameter settings, the power consumption of the PHP may be dominated by the RF transmit power and it may result in a larger power consumption than the FHP.
	    For the cases with 2 BSs and 4 users, implementation of the FHP and the PHP in cooperative networks reduces the required RF transmit power, compared to the case in a non-cooperative network, by $71\%$ and $56\%$, respectively.
	
	\end{abstract}

%\begin{IEEEkeywords}
%	Millimeter Wave, hybrid beamforming, cooperative networks, energy efficiency, user association.
%\end{IEEEkeywords}

 	\section{Introduction}
     5G networks are seen as the next revolution in wireless communications, promising high bandwidth, high energy efficiency, wide coverage, high reliability and low latency \cite{whatwill5gbe}. Particularly, compared to 4G, future networks are expected to support a wide range of use cases, including enhanced mobile broadband with $10^3$ times higher user data rate,  massive machine type communications with $10^4$ times more connected low-cost devices, and critical machine type communications with ultra reliability and low latency in the order of milliseconds \cite{lin20185g}. In order to meet the requirements, different key technologies are currently being considered in 5G, among which network densification \cite{femtocells}, millimeter wave (mmWave) and massive multiple-input multiple-output (MIMO) \cite{massivemimo} are of particular interest.

    Due to the ever-growing mobile data demand, conventional networks operating on the sub-6 GHz spectrum are becoming overcrowded. On the other hand, the mmWave spectrum, in the range of 6-100 GHz, remains largely unused. 
    Therefore, one of the key features of 5G is to utilize the large bandwidth at carrier frequency $>6$ GHz and to provide gigabytes-per-second data rate.
    However, one of the main drawbacks of mmWave frequency signals is the high path loss, which may result in a shorter communication range.
    With the roll out of small BSs, which only cover an area of few hundreds of meters \cite{femtocells, tractable_approach,chao_egc} and the adoption of large antenna arrays that can provide high array gain and directional transmissions, mmWave-enabled networks can become practical.
	Other challenges in mmWave cellular systems include channel modeling and precoder design. Channel measurements have been conducted on various carrier frequencies in various scenarios \cite{mmMAGIC,Rappaport_wideband}, in order to validate simulated data with field measurements and identify the requirements of key system design parameters. As mmWave signals are sensible to blocking, the difference between line-of-sight (LOS) and non-line-of-sight (NLOS) paths need to be considered in the path loss estimation. Environment-dependent path loss exponents and LOS probabilities were proposed in recent studies to improve the accuracy of the path loss models for mmWave communications
     \cite{Bai_CoverageNrate, 3GPP_pathloss, Rajagopal_humanbody, chao_PZF}.

    In order to compensate for the high path loss and maintain a sufficient link budget, precoding with large number of antennas is essential for mmWave BSs to serve multiple users with multiple data streams simultaneously. However, the conventional fully digital precoding (FDP) architecture, which requires a complete radio-frequency (RF) chain and a digital-to-analog converter (DAC) or analog-to-digital converter (ADC) per antenna, may be infeasible as the cost of hardware increases rapidly with the number of antennas. To reduce the cost as well as the complexity of the hardware, the hybrid beamforming architectures, which use less RF chains and DACs/ADCs than the number of antennas, have been proposed \cite{Ayach_SpatiallySparsePrecoding, Rusu_LowComplexityHB, Ni_BlockDiagonalization, Sohrabi_HybridLargeScaleAntenna, Gao_paritiallyconnected, Yu_AlternatingMin, Zhou_StreamAdaptation, Mo_lowadc, Roi_PhaseShifterorSwitch, Lin_EnergyIndoor,He_EnergySubarray, Ge_ComputationPower}. 
    
    Depending on the extent of connections between the RF chains and antennas, the hybrid precoding architecture can be categorized as the fully-connected hybrid precoding (FHP) and the partially-connected hybrid precoding (PHP) architecture.
    The FHP architecture requires each antenna to be connected to all RF chains via phase shifters (PSs) and has been shown to achieve a spectral efficiency close to that of the FDP in a single BS setup\cite{Ayach_SpatiallySparsePrecoding, Rusu_LowComplexityHB,Ni_BlockDiagonalization, Sohrabi_HybridLargeScaleAntenna}. Even though the FHP reduces the power consumption by using less RF chains, it may still need a large number of PSs as the number of antennas grows,
    which poses other challenges such as high PS power consumption, insertion loss, and wiring complexity \cite{Gao_paritiallyconnected}. For this reason, by allowing each RF chain to be connected to a part of the antennas, the PHP architecture is proposed to further reduce the number of PSs, and
    various PHP schemes have been studied to optimize the spectral efficiency \cite{Gao_paritiallyconnected, Yu_AlternatingMin, Zhou_StreamAdaptation} and achieve better energy efficiency than the FHP architecture \cite{Mo_lowadc, Roi_PhaseShifterorSwitch, Lin_EnergyIndoor,He_EnergySubarray, Ge_ComputationPower}. 
    
    In multi-cell scenarios, coordination between BSs is required to alleviate inter-BS interference and reduce service outage. If multi-user channel state information (CSI) is shared among the BSs, power allocation and beam directions can be coordinated for interference management. Furthermore, if user data is available to all BSs, a full cooperation allows a user to receive multiple data streams from different BSs. In this way, the precoders can be jointly designed utilizing the global CSI information of the network.
	For the FDP, many cooperation schemes have been proposed to optimize the system performance in multi-cell networks \cite{Hong_clustering,Jingya_JointPrecodingLoadBalancing, Cheng_CompMixedIntergerConic, Yang_Opportunistic}. 
    Also, coordinated multi-cell systems based on hybrid precoding have been studied in terms of spectral efficiency optimization \cite{Michaloliakos_userbeamselection, Sun_MultiCellHB}. Considering predefined beam patterns and serving each user from only one BS, analog precoders have been jointly selected to maximize the users' data rates \cite{Michaloliakos_userbeamselection}. In \cite{Sun_MultiCellHB}, the performance of interference coordination based on the signal-to-leakage-plus-noise-ratio and regularized zero-forcing hybrid precoding methods were studied in terms of spectral efficiency. In addition, a measurement-campaign-based study for cooperative mmWave BSs is shown to significantly reduce the outage and improve spectral efficiency\cite{Mcartney_Diversity}.

    Despite the advantage of multi-cell cooperation in terms of high spectral efficiency, there is a lack of studies on optimizing the energy efficiency of hybrid-precoding-based cooperative multi-cell mmWave networks. Studying such problems is of interest, because one of the main motivations for hybrid precoding is to leverage large antenna array gains with low cost and, consequently, low power consumption hardware. By reducing the number of RF chains, the number of operations in digital processing are also reduced which further saves energy.
    In order to understand how power savings of specific architectures improve the energy efficiency of mmWave systems, power models that consist of only hardware power consumption are studied by different works. In single-user single-cell mmWave systems, the models that consist of the power consumption of multiple hardware components are proposed in \cite{Abbas_HBADCComparison, Roi_PhaseShifterorSwitch}. Similar models are used to maximize the energy efficiency in multi-user multi-cell mmWave systems \cite{chao_1}.
     
    In order to meet the rate constraints and achieve the objective of minimizing the sum power consumption, 
    cooperative communication that makes the BSs silent in some channel realizations can potentially achieve better interference management and higher useful signal power in each frame \cite{Jingya_JointPrecodingLoadBalancing, Feng_BSsleep}. The BS silence strategy is implemented for FDP and the results show that putting BS into silent mode may be important for load balancing and energy saving \cite{Jingya_JointPrecodingLoadBalancing}. In \cite{Feng_BSsleep}, a review of BS silence strategies for different applications is given.

    In this paper, we study the rate-constrained power efficiency of cooperative mmWave networks using hybrid precoding. The contributions of this paper are as follows. We propose a cooperative hybrid precoding algorithm that minimizes the sum power consumption of the network for FHP and PHP architectures, for both single carrier and OFDM systems. We decouple the hybrid precoding problem into an analog precoding problem with a solution based on equal gain transmission and a digital precoding problem which is in the form of a relaxed semidefinite program subject to per-user rate constraints and per-BS maximum power constraints. For both the FHP and PHP architectures, 
    our proposed hybrid precoding algorithm jointly associates users to the BSs, finds the optimal BS silence strategy with minimum power, and enables us to jointly serve a user by multiple BSs. 
    
    In order to reduce the complexity of our developed cooperative hybrid precoding algorithm, we propose a sub-optimal algorithm, in terms of the BS silence strategy. The proposed sub-optimal algorithm modifies the original objective function to its convex envelope and updates the objective function in each iteration in order to drive the BSs with small RF transmit power into silent mode.
    Also, we present a fairly realistic power model to compare the power consumption of different architectures in a multi-cell and multi-user scenario. We consider a BS silence strategy which minimizes the total power consumption. The power model takes into account both the RF transmit power, which is affected by the power amplifier efficiency and a power loss factor at the BS, as well as the power consumption of hardware components such as the RF chains, the PSs and the DACs. 
    Furthermore, we study the value of BSs cooperation in terms of the power consumption, the probability of infeasible solutions and the probability of joint transmission for FHP and PHP.

	 For single carrier systems, the simulation results are compared among FDP, FHP, PHP and the cases with the sub-optimal algorithm.
	 It verifies that our proposed FHP gives close performance to the FDP, in terms of beam patterns and RF transmit power. Also, we show that if the power consumption is dominated by the RF transmit power, the power consumption of the PHP may be higher than that of the FHP.
	 Furthermore, the simulation results confirm the value of cooperation, as it reduces the sum power consumption of the BSs and the
	 probability of infeasible precoding solutions. For example, when the network changes from 1 BS to 2 BSs with cooperation, for a per-user spectral efficiency of 4 bit/s/Hz and 4 users, the sum RF transmit power is reduced by $71\%$ and $56\%$ for the FHP and the PHP schemes, respectively. For OFDM systems, we show that the energy efficiency of FDP might decrease with more cooperative BSs due to the additional high hardware power consumption.

	 As opposed to \cite{Ayach_SpatiallySparsePrecoding, Rusu_LowComplexityHB, Ni_BlockDiagonalization, Sohrabi_HybridLargeScaleAntenna, Gao_paritiallyconnected, Yu_AlternatingMin, Zhou_StreamAdaptation, Mo_lowadc, Roi_PhaseShifterorSwitch, Lin_EnergyIndoor,He_EnergySubarray, Ge_ComputationPower}, we study the system performance in cooperative mmWave multi-cell scenarios, and consider a fairly realistic model for the power consumption. Moreover, our optimization problem formulation and the proposed hybrid precoding scheme are different from those considering the FDP \cite{Hong_clustering,Jingya_JointPrecodingLoadBalancing, Cheng_CompMixedIntergerConic, Yang_Opportunistic}. In \cite{chao_1}, we performed initial studies on the performance of hybrid precoding schemes. Compared to \cite{chao_1}, our current work presents a cooperative hybrid beamforming algorithm for FHP and PHP and for both  single-carrier and orthogonal frequency-division multiplexing (OFDM) systems. Finally, our results on the effect of cooperation and comparison of different schemes in rate-constrained conditions have not been presented before. The differences in the problem formulation and the power model makes our analytical/simulation results and conclusions completely different from the ones in the literature, e.g., \cite{Ayach_SpatiallySparsePrecoding, Rusu_LowComplexityHB, Ni_BlockDiagonalization, Sohrabi_HybridLargeScaleAntenna, Gao_paritiallyconnected, Yu_AlternatingMin, Zhou_StreamAdaptation, Mo_lowadc, Roi_PhaseShifterorSwitch, Lin_EnergyIndoor,He_EnergySubarray, Ge_ComputationPower, Hong_clustering,Jingya_JointPrecodingLoadBalancing, Cheng_CompMixedIntergerConic, Yang_Opportunistic,Michaloliakos_userbeamselection, Sun_MultiCellHB, Mcartney_Diversity,Abbas_HBADCComparison, chao_1}.

	Notations: We use bold lower-case letters like $\mathbf{d}$ for vectors and upper-case bold letters like $\mathbf{R}$ for matrices. Then, $\mathbf{R}^T$, $\mathbf{R}^H$, $\mathbf{R}^{(i,j)}$ and $||\mathbf{R}||_\mathrm{F}$ denote the transpose, the Hermitian, the $(i,j)$-th entry of $\mathbf{R}$ and the Frobenius norm of $\mathbf{R}$, respectively. Furthermore, $||\mathbf{d}||$ denotes the Euclidean norm and $\Tr(\cdot)$ is the trace of a square matrix. Finally, $\mathbb{C}^{n\times 1}$ represents the set of $n$-tuples of complex numbers represented as column vectors and $\mathbb{C}^{m\times n}$ denotes the set of complex $m \times n$ matrices.
		
	\section{System Model}
	
		\begin{figure}
		\centering
		\begin{subfigure}[b]{\linewidth}
			\centering
			\includegraphics[width=\textwidth]{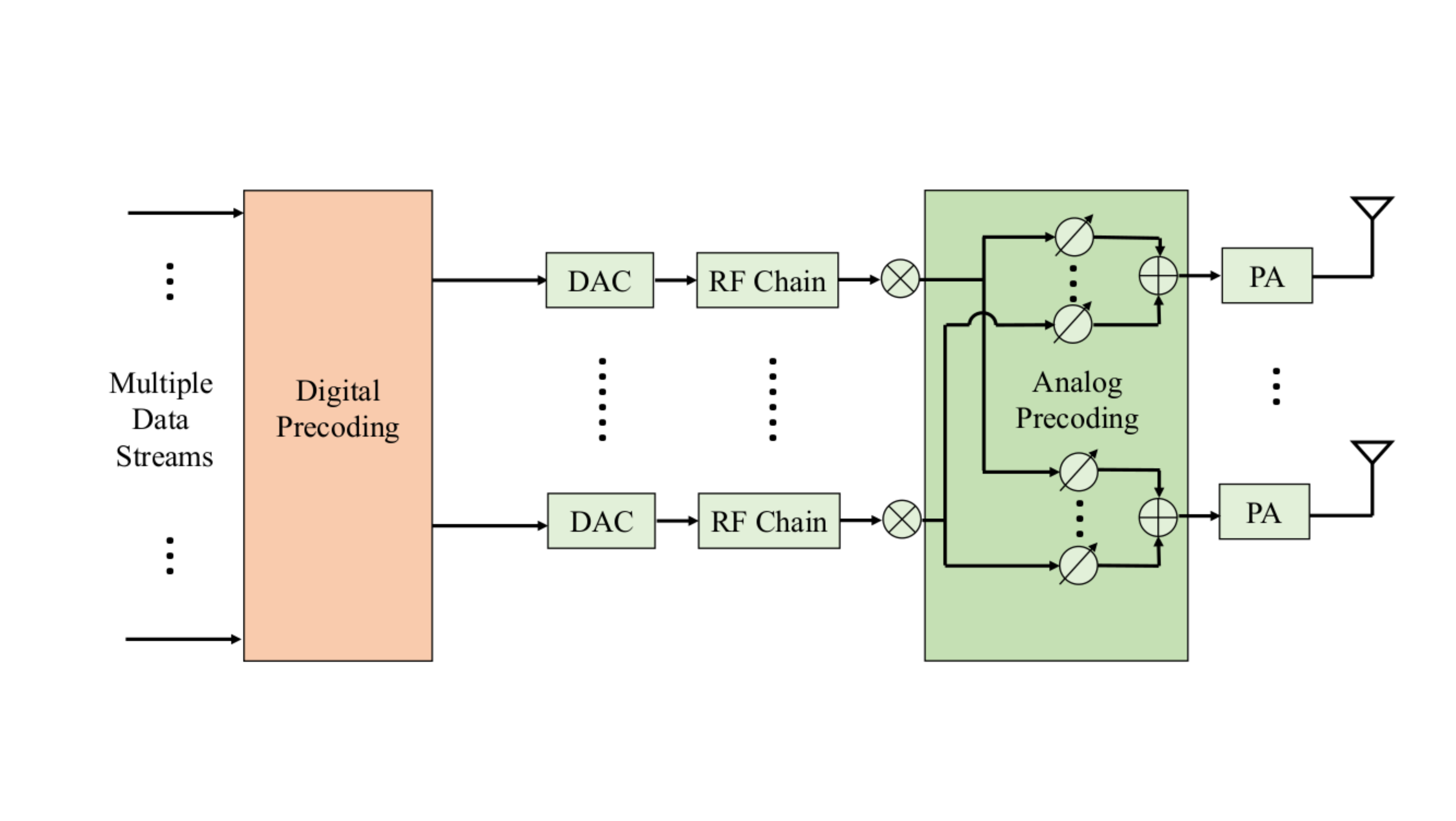}
			\vspace{-1cm}
			\caption{FHP architecture.}
		\end{subfigure}
		\begin{subfigure}[b]{\linewidth}
			\centering
			\includegraphics[width=\textwidth]{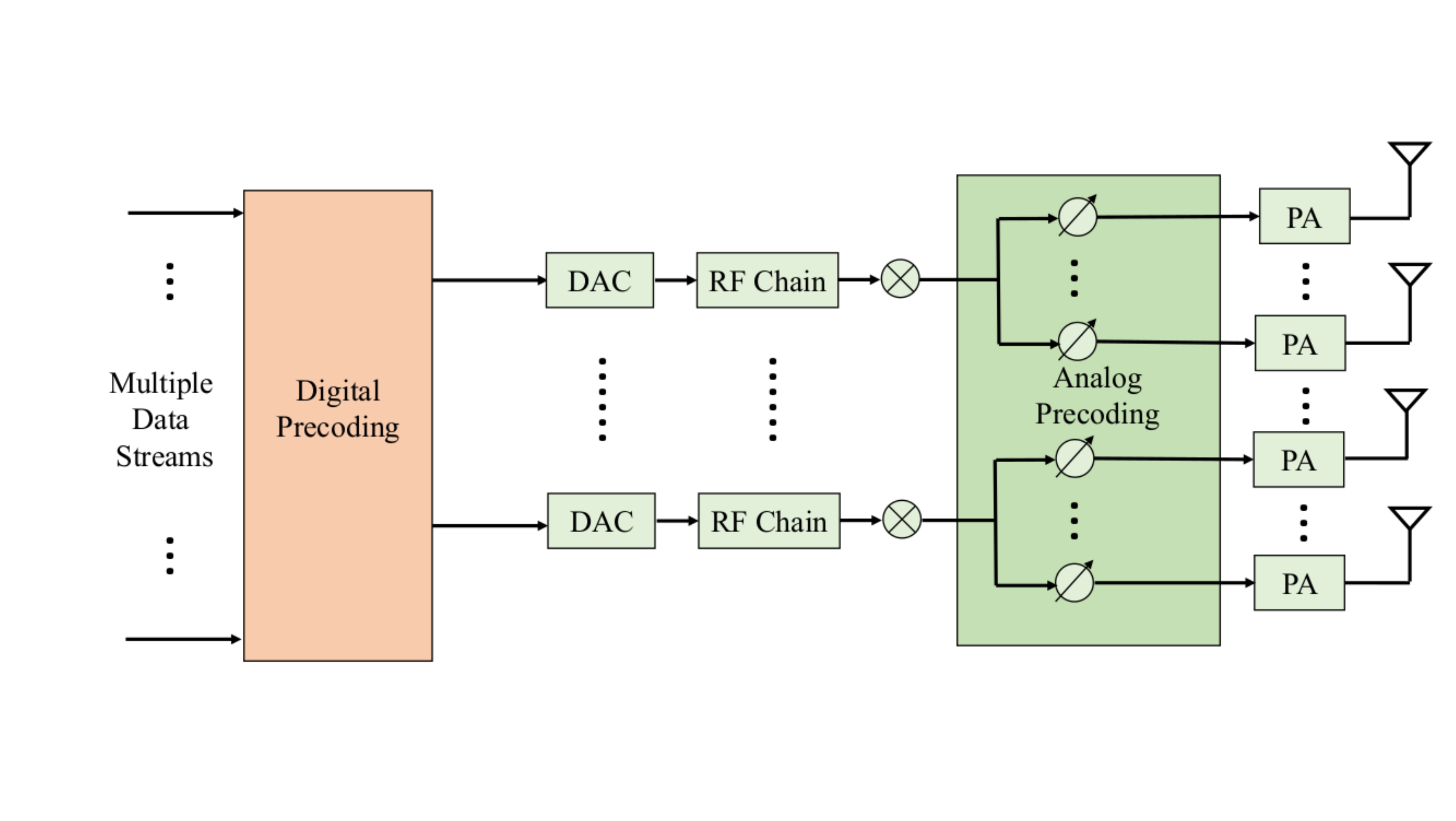}
			\vspace{-1cm}
			\caption{PHP architecture.}
		\end{subfigure}
		\caption{Hybrid precoding architectures.}
		\label{archtecture}
	\end{figure}
	
	We consider a multi-cell multi-user mmWave system with $K$ users and $M$ BSs of, possibly, different types which are characterized by  different transmission power limits and hardware. The number of antennas and the number of RF chains at a BS $m$, $0< m \le M,$ are denoted by $N_m$ and $L_m$, respectively. For tractable analysis, a user $k$, $0 < k \le K,$ is assumed to be equipped with a single antenna or an array with each antenna element receiving beams from a certain angle range. The BSs use hybrid precoding, which satisfies $N_m > L_m \ge K$, to transmit to the users. For $N_m = L_m$, the case of the FDP is considered.

	In the hybrid beamforming architecture, RF chains are interconnected to antennas via PSs.
	Depending on the extent of interconnections and the number of PSs needed, the hybrid precoding architectures can be further categorized into the FHP
	and the PHP architectures. As shown in Fig.~\ref{archtecture}(a),
	the FHP architecture requires each RF chain to be connected to all antennas via PSs. Thus, the main drawback is that the number of PSs grows fast with the number of antennas, leading to high hardware power consumption and complexity. On the other hand, the PHP architecture reduces the number of PSs and the interconnections between the RF chains and the antennas by connecting each RF chain to an antenna subarray, within which each antenna is connected to one PS. For the PHP, we additionally assume that $N_m/L_m, \forall m,$ is an integer such that each RF chain is connected to an antenna subarray and they do not overlap, as shown in Fig.~\ref{archtecture}(b). 
    Hence, the number of PSs required in the FHP is $L_mN_m$ and it is reduced to $N_m$ in the PHP. For the FDP, because $N_m = L_m$, PSs are not needed to connect the antennas and RF chains.

	The transmit symbol $x_{k,m}$ to user $k$ at BS $m$ is first precoded by a baseband digital precoder $\mathbf{d}_{k,m} \in \mathbb{C}^{L_m\times 1}$ then followed by an analog precoder $\mathbf{R}_{m} \in \mathbb{C}^{N_m\times L_m}$, such that the precoded signals at particular angles have strong power while causing less interference to other users.
	The digital precoder has full control over both the amplitude and the phase of the signal, while the analog precoder is enabled by PSs and can only change the phase of the signal. 
	For the FHP, the entries in the analog precoder have the constant amplitude constraint $\left|\mathbf{R}_{m}^{(i,j)}\right|=\frac{1}{\sqrt{L_mN_m}}, \forall m$. For the PHP, each RF chain is connected to part of the PSs, therefore the analog precoder has the form of a block diagonal matrix given by $\mathbf{R}_m = \mathrm{diag}[\hat{\mathbf{r}}_{1,m}, \cdots, \hat{\mathbf{r}}_{L_m,m}]$.
	Here, $\hat{\mathbf{r}}_{i,m}\in \mathbb{C}^{{N_m}/{L_m}\times 1}, 1\le i\le L_m,$ with constraints $\left|\hat{\mathbf{r}}_{i,m}^{(j)}\right|= 1/\sqrt{N_m}, \forall m$.

    \subsection{Channel Model}
    Since large antenna arrays and directive transmissions make the mmWave multi-path channel sparser than the lower frequency channel, we consider a channel model that includes a few dominant signal clusters where each cluster contains several multi-path components. Assuming a quasi-static condition, which is a good channel model for the cases with slow-moving or static users, and single-carrier channel model, the narrow-band channel between BS $m$ and user $k$ is given by \cite{Ayach_SpatiallySparsePrecoding, Rusu_LowComplexityHB,Ni_BlockDiagonalization,Sohrabi_HybridLargeScaleAntenna, Gao_paritiallyconnected, Yu_AlternatingMin}
    % Channel model
    \ieb
	    \mathbf{h}_{k,m} = \sqrt{\frac{\rho_{k,m}N_m}{N_{\mathrm{cl}}N_{\mathrm{ray}}}}
	    \sum_{i=1}^{N_{\mathrm{cl}}}\sum_{l=1}^{N_{\mathrm{ray}}}\alpha_{i,l}\mathbf{a}^t_m(\theta^t_{i,l}),
    \label{nbchannel}
    \ien
    where $N_{\mathrm{cl}}$ is the number of clusters, $N_{\mathrm{ray}}$ is the number of paths within a cluster, $\alpha_{i,l}\sim \mathcal{CN}(0,1)$ is the gain of the $l$-th path in the $i$-th cluster and $\mathbf{a}^t_m(\theta^t_{i,l})$ is the normalized transmit antenna array response vector. For an $N_m$-element uniform linear array, it is given by
    \ieb
    \mathbf{a}^t_m(\theta^t_{i,l})=\frac{1}{\sqrt{N_m}}\left[1, e^{jkd\sin(\theta^t_{i,l})}, \cdot\cdot\cdot, e^{jkd(N_m-1)\sin(\theta^t_{i,l})}\right]^T, \nonumber \\
     \label{apattern1}
    \ien
    where $\theta^t_{i,l}$ is the angle of departure (AOD), $k=2\pi/\Lambda$, $\Lambda$ is the wavelength and $d=\Lambda/2$ is the antenna spacing. Also, $\theta^t_{i,l}$ is assumed to follow a truncated Laplace distribution with mean cluster angle $\bar{\theta}_i \sim \mathcal{U} (\theta_i^{\text{min}}, \theta_i^{\text{max}})$ and angular spread $\sigma_{\theta_i}$, which assumes the transmitter uses sector transmissions.
    
    The path loss in mmWave channels differs greatly depending on the LOS and NLOS environment \cite{Samimi_ProbabilisticPL,Bai_CoverageNrate, Mcartney_Diversity}. In order to incorporate both the LOS and NLOS models together, the path loss of LOS or NLOS transmissions is determined by a LOS probability as a function of the transmission distance. In this way, the path loss between BS $m$ and user $k$ is given by 
    \ieb
    \rho_{k,m}=\mathbb{I}(p_L(\hat{d}))\mathrm{PL}^{-1}_{\textrm{LOS}}
    +\left(1-\mathbb{I}(p_L(\hat{d})) \right)\mathrm{PL}^{-1}_{\textrm{NLOS}},
    \label{pathloss}
    \ien
    where $\hat{d}$ is the distance and $\mathbb{I}(p_L(\hat{d}))$ is a Bernoulli random variable. The LOS probability is given by $p_L(\hat{d})=e^{-\beta \hat{d}}$ \cite{Bai_Blockage}, where the exponentially decaying probability models the fact that the probability of LOS decreases with distance and $\beta$ models the average blockage density that causes the NLOS condition. The path loss follows the close-in free space reference distance model and is of the form \cite{Samimi_ProbabilisticPL, Mcartney_Diversity}
    \ieb
    &\mathrm{PL}_{\mathrm{LOS/NLOS}}[\mathrm{dB}] &= 20\log_{10}\left(\frac{4\pi}{\lambda}\right) + 10 \bar{n}_{\mathrm{LOS/NLOS}} \nonumber\\
    &&\times\log_{10}(\hat{d}) 
    + X_{\mathrm{LOS/NLOS}}, \hat{d}\ge 1~\mathrm{m},
    \ien
    where $\bar{n}_{\mathrm{LOS/NLOS}}$ and $X_{\mathrm{LOS/NLOS}}$ are the LOS and NLOS dependent path loss exponent and log-normal distributed shadowing  with standard deviation $\sigma_{\mathrm{LOS/NLOS}}$, respectively. 

	\subsection{Spectral Efficiency}	
	Considering quasi-static channels known by the BSs, the composite signal, which is the sum of all signals sent from the BSs,  received by user $k$ can be written as
	\ieb
	y_k &=& \sum_{m=1}^M \mathbf{h}^H_{k,m} \mathbf{w}_{k,m}x_{k,m} + \sum_{m=1}^{M}\sum_{k'\ne k,k'=1}^K\mathbf{h}^H_{k,m}
	\mathbf{w}_{k',m}x_{k',m} \nonumber \\	
	&&+ n_k.
	\label{signal_y}
	\ien
	Here, $\mathbf{w}_{k,m}=\mathbf{R}_{m}\mathbf{d}_{k,m}$ is the 
	overall precoder, 
	$\mathbf{h}_{k,m} \in \mathbb{C}^{N_m \times 1} $ is the channel vector, $x_{k,m}$ is the data symbol with $\mathbb{E}[x_{k,m}x_{k,m}^H]=1$, $n_k\sim \mathcal{CN}(0, \sigma_k^2) $ is the complex Gaussian noise with variance $\sigma_k^2$, and the second term is the interference.
	In the case of FDB, the precoder is reduced to $\mathbf{w}_{k,m}=\mathbf{d}_{k,m}$ by setting $\mathbf{R}_m = \mathbf{I}$ and $N_m=L_m$.

	To achieve cooperative transmissions, it is often required that the channel state information (CSI) is shared between the BSs \cite{Gesbert_multicell,Jingya_JointPrecodingLoadBalancing, Larsson_JointPowerAllocation}, in order to make joint decisions on different operations such as user scheduling, precoding, combining, power control, etc.
	We assume that a user $k$ is able to receive useful signals from multiple BSs, where the data symbols $x_{k,m}$ from different BSs are assumed to be mutually independent \cite{Jingya_JointPrecodingLoadBalancing, Larsson_JointPowerAllocation}. The user $k$ implements successive interference cancellation 
	to decode the data streams sequentially, i.e., the stronger signals are decoded first and are then subtracted from the received signals to decode the weaker signals.
	We assume that there is a central node which gathers the channel conditions of all BSs-user pairs and jointly performs the precoding calculations. In this way, each BS only needs to share its own channel, hence the CSI exchange overhead can be maintained low. Moreover, BSs timing and frame synchronization may not be an major issue if these BSs are connected via wired backhaul.

	Assuming Gaussian signaling, the achievable spectral efficiency for user $k$ is given by
		\ieb
		\label{SINR_k}
		\Gamma_k=\log_2 \left(1+\frac{\sum_{m=1}^{M}|\mathbf{h}_{k,m}^H\mathbf{w}_{k,m}|^2}{I_k+\sigma_k^2}\right),
		\ien
		where $I_k=\sum_{m=1}^{M}\sum_{k'\ne k,k'=1}^{K}\mathbf{w}_{k',m}^H\mathbf{h}_{k,m}\mathbf{h}_{k,m}^H\mathbf{w}_{k',m}$ is the interference power.
	
	\subsection{Power Consumption Model}
	 In this subsection, we present the power consumption model for the FDP, the FHP, and the PHP. The model includes several main power consuming hardware components, which helps us to have fair comparison of the power consumption among different precoding architectures.
	 
	 We consider that the power consumption of a BS consists of the hardware power consumption and the RF transmit power. The RF transmit power at BS $m$ is given by
	 \ieb
	 \label{trans_power}
	 P^{\mathrm{tx}}_{m} =
	 \sum_{k=1}^{K}||\mathbf{w}_{k,m}||^2,
	 \ien
	and the hardware power consumption of phase shifters, RF chains and the DACs is given by
	 	\ieb
	 \label{hw_power}
	 P^{\mathrm{hw}}_m = \frac{ N_m^{\mathrm{PS}}P_{\mathrm{PS}} + L_m\left(P_{\mathrm{DAC}}+P_{\mathrm{RF}}\right)}{1-\Delta_m}.
	 \ien  
	 Here, $\Delta_m$ is a power loss factor which accounts for the extra power dissipated at various non-transmission related parts such as power supply loss and active cooling \cite{EARTH}. Also, $P_{\mathrm{PS}}, P_{\mathrm{DAC}}$ and $P_{\mathrm{RF}}$ denote the power consumption of the PSs, DACs and RF chains, respectively, and $N_m^{\mathrm{PS}}$ is the number of PSs. The number of PSs needed by each architecture is given by the FDP: $N_m^{\mathrm{PS}}= 0, L_m=N_m$, the FHP: $N_m^{\mathrm{PS}}= L_mN_m$, and the PHP: $N_m^{\mathrm{PS}}= N_m$.

	To further reduce the power consumption, we assume that, when there are no users associated to a BS, this BS can be put into silent mode by deactivating some of the hardware. At the silent mode, a BS consumes a proportion of the hardware power, which is denoted by $aP^{\mathrm{hw}}_m$, where $a \in [0,1]$ is the silent mode scalar whose magnitude depends on the number of hardware components deactivated during the silent mode and affects the delay needed to reactive the BS \cite{Debaillie_PowerModel}. Including the silent mode, the total power consumption of BS $m$ is given by
	\ieb
	P_m=
	\begin{cases}
		\eta_m'P^{\mathrm{tx}}_{m} +  P^{\mathrm{hw}}_m,  ~~~~\mathrm{active,}\\
		 aP^{\mathrm{hw}}_m,~~~~~~~~~~ P_{m}^\mathrm{tx}=0,~\mathrm{silent},
	\end{cases}
     \label{powermodel}
	\ien
	 where $\eta_m'=\frac{1}{\eta_m(1-\Delta_m)}$ and $\eta_m$ denotes the power amplifier efficiency.
    
	\section{Cooperative Hybrid Beamforming}
    Considering the hardware power consumption and the silent mode, for each of the considered precoding architectures, i.e., FDP, FHP and PHP, 	
   	our objective is to find the analog and digital precoders ($\mathbf{R}_{m}, \mathbf{d}_{k,m}, \forall m, k$) such that the sum power consumption is minimized.
   	 The problem can be summarized as
    		\ieb
    		\label{objective}
    		&&\mathcal{P}:\min_{\{\mathbf{R}_{m},\mathbf{d}_{k,m}\}}\sum_{m=1}^{M} b_mP_m \\
    		&&\mathrm{s.t.}~~~ \Gamma_k\ge \tau_k, ~~~~~ \forall k \label{rate_constraint}\\
    		\label{power_constraint}
    		&&\sum_{k=1}^{K}||\mathbf{R}_{m}\mathbf{d}_{k,m}||^2 \le P_{\max,m},~~~~~ \forall m\\
    		\label{analog_constraint}
    		&&\left|\mathbf{R}_{m}^{(i,j)}\right|=\frac{1}{\sqrt{N_mL_m}}, ~~~~~ \forall m,i,j,~~~~~\mathrm{for}~\mathrm{FHP}, \label{FHP_constraint}\\
    		&&\mathbf{R}_m = \mathrm{diag}[\hat{\mathbf{r}}_{1,m}, \cdots, \hat{\mathbf{r}}_{L_m,m}],\nonumber\\ &&\left|\hat{\mathbf{r}}_{i,m}^{(j)}\right|= 1/\sqrt{N_m}, ~~~~~ \forall m,j, ~~~~~\mathrm{for}~\mathrm{PHP}.
    		 \label{PHP_constraint}
    		\ien
    		where $b_m$ is the weighting parameter for balancing the load of BS $m$,
    		\eqref{rate_constraint} is the minimum spectral efficiency constraint for each user with target spectral efficiency  $\tau_k$, and \eqref{power_constraint} is the maximum RF transmit power constraint on each BS. For the hybrid precoding architectures, the additional constraints \eqref{FHP_constraint} and \eqref{PHP_constraint} need to be considered for the FHP and PHP, respectively.
    		The solution to problem $\mathcal{P}$ gives the analog precoder for each BS and the digital precoder for each BS-user pair. It also reflects the user association strategy, as the set of associated users of BS $m$ is given by 
	    	 $\mathcal{K}_m=\{k| 0 < k\le K, ||\mathbf{R}_{m}\mathbf{d}_{k,m}||^2>0\},$
	    	and the set of serving BSs of UE $k$ is given by
	    	 $ \mathcal{M}_k=\{m| 0<m\le M, ||\mathbf{R}_{m}\mathbf{d}_{k,m}||^2>0\}.$
    		
    		Solving problem $\mathcal{P}$ is challenging, because it is a non-convex optimization problem due to the analog precoder constraints \eqref{FHP_constraint} and \eqref{PHP_constraint}. Moreover, the number of design parameters involved is large due to the fact that we have to jointly design precoders for each user-BS pair. Therefore, for the single-BS setup, the hybrid precoding optimization is often solved by sub-optimal methods that decouple the analog and digital precoding processes and use 
    		iterative updates over the analog and digital precoders to approach the performance of the FDP \cite{Ayach_SpatiallySparsePrecoding, Ni_BlockDiagonalization, Sohrabi_HybridLargeScaleAntenna, Gao_paritiallyconnected, Rusu_LowComplexityHB,Yu_AlternatingMin}. 
    		However, such iterative optimization methods may not be suitable for cooperative multi-cell and multi-user systems due to the fairly large amount of design parameters and feedback from all BSs, therefore, causing high computation complexity and delay. In order to make the problem tractable, we propose a sub-optimal solution by decoupling the optimization problem $\mathcal{P}$ into an analog precoding problem, which only depends on the channel information, and a digital precoding problem that minimizes the sum power consumption \eqref{objective} conditioned on the analog precoders. The method, as shown in Section IV, achieves close spectral efficiency for a given RF transmit power compared to FDP.
       \subsection{Analog Precoding}
       
       In this subsection, we show how to obtain the analog precoders.  In order to decouple the analog and digital precoders, we cast the analog precoding problem as an equal gain transmission problem which is independent of the digital precoders.
    	
    	For FHP, defining the array gain between user $k$ and BS $m$ as  $g_{m,k}=|\mathbf{h}_{k,m}^H\mathbf{R}_m\mathbf{d}_{k,m}|^2$, using Cauchi-Schwarz inequality, we have
    	\ieb
    	& |\mathbf{h}_{k,m}^H\mathbf{R}_m\mathbf{d}_{k,m}|^2 \le& \left(|\mathbf{h}_{k,m}^H\mathbf{r}_{k,m}|^2+||\mathbf{h}_{k,m}^H(\mathbf{R}_{m})_{-k}||_{\textrm{F}}^2\right)\nonumber \\
    	&&\times  ||\mathbf{d}_{k,m}||^2,
    	\label{upperbound}
    	 \ien
    	where $\mathbf{r}_{k,m}$ is the $k$-th column of $\mathbf{R}_m$ and $(\mathbf{R}_{m})_{-k}$	
    	 is a matrix after removing the $k$-th column from $\mathbf{R}_{m}$. 
    	 Treating $||\mathbf{h}_{k,m}^H(\mathbf{R}_{m})_{-k}||_{\textrm{F}}^2||\mathbf{d}_{k,m}||^2$ as the interference power from other users, we only focus on maximizing the useful signal power $|\mathbf{h}_{k,m}^H\mathbf{r}_{k,m}|^2||\mathbf{d}_{k,m}||^2$ without considering the interference management. Hence, conditioned on the digital precoders, the analog precoding problem is given by
    	\ieb
    	&&\max_{\mathbf{r}_{k,m}}\left|\mathbf{h}^H_{k,m}\mathbf{r}_{k,m}\right|^2 \label{EGT_problem} ~~~~~ \\
    	&&\mathrm{s.t.}\left|\mathbf{r}_{k,m}^{(i)}\right|=\frac{1}{\sqrt{L_mN_m}}, ~~~~~ 0< i\le N_m.
    	\label{EGT_probcon}
    	\ien
	    The optimization problem \eqref{EGT_problem} and \eqref{EGT_probcon}	is equivalent to the equal gain transmission problem \cite{Love_EqualGainTransmission}
    	 which has the analytical solution given by
    	\ieb
    		\label{Analog_sol1}
    	&&\mathbf{r}_{k,m}^{(i)}=\frac{1}{\sqrt{L_mN_m}}e^{j\left(\xi+ \angle\mathbf{h}_{k,m}^{(i)}\right)}, 
    	\ien
    	where $\xi \in (0,2\pi]$ is an arbitrary phase and $\angle\mathbf{h}_{k,m}^{(i)}$ is the phase angle of $\mathbf{h}_{k,m}^{(i)}$.
    	We notice that the equal gain transmission requires $K=L_m$, i.e., the number of serving users is equal to the number of RF chains at BS $m$. 
    	Here, we assume that each BS uses analog precoding to serve all users, 
    	however, after the digital precoding step, a BS may not transmit to all users, as the final users associated to BS $m$ will be determined by both precoders according to $\mathcal{K}_m=\{k| 0< k\le K, ||\mathbf{R}_{m}\mathbf{d}_{k,m}||^2>0\}.$ 
    	
    	For the PHP, we additionally require that $\mathbf{R}_m = \mathrm{diag}[\hat{\mathbf{r}}_{1,m}, \cdots, \hat{\mathbf{r}}_{L_m,m}]$ with $\left|\hat{\mathbf{r}}_{i,m}^{(j)}\right|= 1/\sqrt{L_m}$. Denoting the $k$-th column of $\mathbf{R}_m$ by 
        $\mathbf{r}_{k,m}$, following the same upper bound maximization steps, the analog precoding problem is given by
    	\ieb
    	&&\max_{\mathbf{r}_{k,m}}\left|\hat{\mathbf{h}}^H_{k,m}\mathbf{r}_{k,m}\right|^2 ~~~~~ \label{EGT_prob}\\
    	&&\mathrm{s.t.}\left|\mathbf{r}_{k,m}^{(i)}\right|=\frac{1}{\sqrt{N_m}}, ~~~~~ \frac{(k-1)N_m}{L_{m}}< i\le \frac{kN_m}{L_m}, 
    	\label{EGT_prob1}
    	\ien
    	where $\hat{\mathbf{h}}^H_{k,m}= \mathbf{h}^H_{k,m} \mathbf{G}_{k,m}$ and
    	\ieb
    	 \mathbf{G}_{k,m} &=& \mathrm{diag}\Big(\mathbf{0}_{\frac{(k-1)N_m}{L_m}\times \frac{(k-1)N_m}{L_m}}, \mathbf{I}_{\frac{N_m}{L_m}
    	 	\times\frac{N_m}{L_m}}, \nonumber \\    	
     	&&\mathbf{0}_{\frac{(L_m-k)N_m}{L_m}\times\frac{(L_m-k)N_m}{L_m}}\Big).
     	\label{RF_G}
    	\ien    	
    	The analog precoding problem \eqref{EGT_prob} and \eqref{EGT_prob1} has the same form as \eqref{EGT_problem} and \eqref{EGT_probcon}, therefore, the analytical solution is given by
    		\ieb
    	&&\hat {\mathbf{r}}_{k,m}^{(i)}=\frac{1}{\sqrt{N_m}}e^{j\left(\xi+ \angle \hat{\mathbf{h}}_{k,m}^{((k-1)N_m/L_m+i)}\right)}, 0<i \le N_m/L_m. 
    	\nonumber \\
    	\label{Analog_sol2}
    	\ien
    	
       Although our analog precoding solution based on the equal gain transmission neglects the interference power and digital precoding, it maximizes the signal-to-noise ratio in a single-BS analog beamforming system.  
       The analog precoders only require the channel phase information for each user and has low complexity. The interference coordination and the overall array gain enhancement will be addressed by the digital precoding. As shown in our simulations, for a given RF transmit power, this low complexity analog precoding solution and the proposed non-iterative optimization approach can achieve close spectral efficiency compared to FDP.
    	\subsection{Digital Precoding}
        \label{secdp}
        To achieve our objective in minimizing the sum power consumption as shown in Problem $\mathcal{P}$, next, we reformulate it to a convex semidefinite program conditioned on the analog precoders and obtain the sub-optimal digital precoders in terms of the sum power consumption.

        We first rewrite the objective function \eqref{objective} in its quadratic form. Define $\mathbf{D}_k = \mathbf{d}_k\mathbf{d}_k^H$ with $\mathbf{d}_k\triangleq \left[\mathbf{d}^T_{k,1},..., \mathbf{d}^T_{k,M} \right]^T$  and a block diagonal matrix
		\ieb
		\hat{\mathbf{R}}=\mathrm{diag}\left(b_1\eta'_1\mathbf{R}_1^H\mathbf{R}_1,\cdots,b_M\eta'_M\mathbf{R}_M^H\mathbf{R}_M\right),\label{lemma1_defM}
		\label{rdy}
		\ien
		the sum power consumption of all BSs can be written as 
		\ieb
		\label{objective_qua}
		\sum_{k=1}^{K} \Tr\left( 	\hat{\mathbf{R}} \mathbf{D}_k(\mathbf{z}_i)\right) +\kappa(\mathbf{z}_i),
		\ien
		where $\kappa(\mathbf{z}_i) = \sum_{m=1}^{M} b_m(z_m+a(1-z_m))P_{m}^\mathrm{hw}$
		and $z_m\in \{0,1\}$ is a silent mode indicator for BS $m$. If $z_m=1$, the BS $m$ is in the active mode, otherwise it is in the silent mode. Here, $\mathbf{z}_i$ is one unique silent mode indicator vector out of the $2^M-1$ possible BS silent mode combinations. 
		
		Next, we express the spectral efficiency and the peak power constraints in more compact forms. Define the block diagonal matrices
		\ieb
		\hat{\mathbf{H}}_k= \mathrm{diag}\left(\mathbf{R}_{1}^H\mathbf{h}_{k,1}\mathbf{h}_{k,1}^H\mathbf{R}_{1},\cdots,\mathbf{R}_{M}^H\mathbf{h}_{k,M}\mathbf{h}_{k,M}^H\mathbf{R}_{M}\right), \nonumber\\
		 \label{lemma1_defH}
		\ien
		where $\mathbf{h}_{k,m}^H\mathbf{R}_{m}$ is the effective channel of user $k$ and BS $m$ after analog precoding. Thus, \eqref{SINR_k} can be rewritten as 
		\ieb
		\label{rate_qua}
		\Gamma_k =\log_2\left(1+ \frac{\mathbf{d}_k^H \hat{\mathbf{H}}_k\mathbf{d}_k}{\sum_{\substack{k'\ne k , k'=1}}^{K}\mathbf{d}_{k'}^H \hat{\mathbf{H}}_{k}\mathbf{d}_{k'}+\sigma_k^2} \right).
		\ien
	   To satisfy a minimum spectral efficiency $\tau_k$, \eqref{rate_qua} can be transformed to the following quadratic inequalities by using the cyclic property of the trace operation:
		\ieb
		\text{Tr}\left(\mathbf{D}_k \hat{\mathbf{H}}_k\right)-(2^{\tau_k}-1)\sum_{\mathclap{k'\ne k, k'=1}}^{K}\text{Tr}(\mathbf{D}_{k'} \hat{\mathbf{H}}_{k})\ge(2^{\tau_k}-1)\sigma_k^2, ~\forall k. \nonumber \\
		\label{lemma1_temp1}
		\ien
		In order to rewrite the maximum transmit power constraints to the quadratic form, we first define
		\ieb
		\mathbf{Q}_{i,m} = 
		\begin{cases}
			\mathbf{R}_m^H\mathbf{R}_m,& \mathrm{if} ~~ i=m \\
			\mathbf{0}_{L_m\times L_m}, & \mathrm{otherwise}
		\end{cases}
		\ien
		and $\mathbf{Q}_m=\mathrm{diag}\left(\mathbf{Q}_{1,m},\cdots,\mathbf{Q}_{M,m}\right)\label{lemma1_defQ}$, then \eqref{power_constraint} can be expressed as 
		\ieb
		\sum_{k=1}^{K} \Tr\left(\mathbf{Q}_m\mathbf{D}_k\right) \le z_m P_{\mathrm{max},m}, ~~~~~ \forall m.
		\label{maxpower_qua}
		\ien
		
		Combining \eqref{objective_qua}, \eqref{lemma1_temp1}, and \eqref{maxpower_qua}, for given $\mathbf{R}_m$ and $\mathbf{z}_i$, we have the following digital precoder optimization problem 
		\ieb
		\label{lemma1_objective}
		&&\mathcal{P}_1 : \min_{\{\mathbf{D}_k\}} \sum_{k=1}^{K} \Tr\left( 	\hat{\mathbf{R}} \mathbf{D}_k\right) + \kappa(\mathbf{z}_i) \\
		&\text{s.t.}~~~&\Tr\left(\mathbf{D}_k \hat{\mathbf{H}}_k\right)\ge(2^{\tau_k}-1)\sum_{\mathclap{k'\ne k, k'=1}}^{K}\Tr\left(\mathbf{D}_{k'} \hat{\mathbf{H}}_k\right)\nonumber \\
		&&+(2^{\tau_k}-1)\sigma_k^2, ~~~~ \forall k,\\
		\label{lemma1_power}
		&&\sum_{k=1}^{K} \Tr\left(\mathbf{Q}_m\mathbf{D}_k\right) \le z_mP_{\mathrm{max},m}, ~~~~~ \forall m.
		\ien
		Problem $\mathcal{P}_1$ is a convex semidefinite problem if we apply the semidefinite relaxation by replacing the constraint $\text{rank}\left(\mathbf{D}_k\right)=1$ with $\mathbf{D}_k\ge0$, where $\mathbf{D}_k\ge0$ denotes positive semidefinite matrices. Then, it can be solved efficiently using convex optimization tools and a rank-1 solution always exists given that the solution is feasible \cite{Jingya_JointPrecodingLoadBalancing}. Problem $\mathcal{P}_1$ is a two-layer inner-outer optimization where we solve for $D_k$ for fixed $\mathbf{z}_i$. The solution that minimizes the sum power
		consumption considering the BS silent/active modes is then obtained by solving $\mathcal{P}_1$ for all possible $\mathbf{z}_i$. Moreover, the user association is found by checking the user-BS pairs with non-zero $||\mathbf{R}_m\mathbf{d}_{k,m}||^2.$ 
		
		The digital precoding problem is conditioned on the analog precoders given by \eqref{Analog_sol1} or \eqref{Analog_sol2} and is independent of the hybrid precoding architectures.
		Although the proposed analog precoding does not guarantee the spectral efficiency and may cause some interference, the phase and amplitude are further adjusted via the digital precoding. The final sub-optimal solution in terms of the sum power consumption is based on the effective channel after analog precoding and ensures that the spectral efficiency and power constraints for all users and BSs are satisfied.

		\begin{algorithm}[t]
			\caption{Hybrid Precoding for Cooperative MmWave Networks}
			\label{algo_EGT}
			\begin{algorithmic}[1]
				\Require{$\mathbf{h}_{k,m}, \forall k,m$, $\tau_k,\forall k$, $P_{\mathrm{max},m} \forall m$.}
			%	\Statex
				\State  $\mathbf{R}_m= \mathbf{0}_{N_m\times L_m}$ 
				\For{$m \gets 1~\textrm{to}~M$}
				\For{$k \gets 1~\textrm{to}~K$}
				\State $\mathbf{r}^{(i)}_{k,m}=\frac{1}{\sqrt{L_mN_m}}e^{\left(\xi+\angle\mathbf{h}_{k,m}^{(i)}\right)}$, for FHP
				\State
				$\hat{\mathbf{r}}_{k,m}^{(i)}=\frac{1}{\sqrt{N_m}}e^{j\left(\xi+ \angle \hat{\mathbf{h}}_{k,m}^{((k-1)N_m/L_m+i)}\right)}$, for PHP
				\EndFor
				\EndFor
				\For{$i \gets 1~\textrm{to}~2^M-1$}
				\State Solve $\mathcal{P}_1$ for fixed $\mathbf{R}_m$ and $\mathbf{z}_i$, obtain $\mathbf{D}_k^*(\mathbf{z}_i)$. 
				\EndFor
				\State
				$o= \argmin_{i} \sum_{k=1}^{K} \Tr\left( 	\hat{\mathbf{R}} \mathbf{D}_k^*(\mathbf{z}_i)\right) + \kappa(\mathbf{z}_i)$, obtain the optimal silent mode indicator $\mathbf{z}_o$,
				\State Minimum sum RF transmit power  $P_{\mathrm{tx}}^*=\sum_{m=1}^{M}\sum_{k=1}^{K} \Tr\left( 	\mathbf{Q}_m \mathbf{D}_k^*(\mathbf{z}_o)\right)$ ,
				\State Minimum sum power $ P^* =\sum_{k=1}^{K} \Tr\left( 	\hat{\mathbf{R}} \mathbf{D}_k^*(\mathbf{z}_o)\right)+ \kappa(\mathbf{z}_o).$ 
			%	\State Let $\mathbf{D}_{k}=\sigma_1\mathbf{u}_1\mathbf{v}_1^H$ be the singular value decomposition, then $\mathbf{d}_k = \sqrt{\sigma_1}\mathbf{u}_1$.
			\end{algorithmic}
		\end{algorithm}

        The overall proposed hybrid precoding for cooperative mmWave networks is described as follows and in Algorithm 1. In the algorithm, $\mathbf{z}_\text{o},$ $P_\text{tx}^*$ and $P^*$ denote the optimal BS silent mode indicator in terms of minimizing the sum power, the sum RF transmit power and the sum power consumption for $\mathbf{z}_\text{o}$, respectively.
		We first find $\mathbf{R}_m$ for all BSs based on \eqref{Analog_sol1} or \eqref{Analog_sol2}, depending on the considered precoding setup. Conditioned on $\mathbf{R}_m$ and $\mathbf{z}_i$, the relaxed semidefinite program in $\mathcal{P}_1$ can be solved efficiently via convex optimization tools. The optimal BS silent/active mode pattern $\mathbf{z}_{\mathrm{o}}$ is given by choosing $\mathbf{z}_i$ with the minimum sum power consumption. Then, the digital precoder based on $\mathbf{D}_k^*(\mathbf{z}_{\mathrm{o}})$ minimizes the sum power consumption.
		The solution $\mathbf{D}_k^*(\mathbf{z}_{\mathrm{o}})$ is rank-1 and the stacked digital precoders $\mathbf{d}_k$ are readily available via singular value decomposition. 

		The complexity of Algorithm 1 consists of the complexity of the analog precoder operations and the complexity of the quadratic programing. For FDP, there is only quadratic programing, therefore the time complexity of FDP is $\mathcal{O}\left((2^M-1)( K\sum_{m=1}^{M}N_m)^3 \right)$. For FHP, the time complexity is a sum of the complexity of the analog precoders and the complexity of the quadratic programing. Therefore, it is given by $\sum_{m=1}^{M}L_mN_m + \mathcal{O}\left((2^M-1)( K\sum_{m=1}^{M}L_m)^3 \right)$. Finally, for PHP, the complexity is given by  $\sum_{m=1}^{M}N_m + \mathcal{O}\left((2^M-1)( K\sum_{m=1}^{M}L_m)^3 \right)$.	Since we focus on a network with a small $M$, the complexity is not problematic.

		\subsection{Lagrangian Analysis}
		In order to gain insights of the cooperative transmissions and the user associations, in this subsection, we analyze the Lagrangian dual problem of the optimization problem and provide conditions for the user association strategy. 
		
		Conditioned on the analog precoders and all BSs being active, the Lagrangian of the optimization problem $\mathcal{P}_1$ is given by
		\ieb
		&&\mathcal{L}(\mathbf{D}_k, \lambda_k, \mu_k)
		 = \sum_{k=1}^{K} \Tr\left( 	\hat{\mathbf{R}}\mathbf{D}_k\right)+\kappa(\mathbf{z})+\sum_{k=1}^{K}\lambda_k\nonumber \\
		 && \times  \Bigg((2^{\tau_k}-1)\Big( \sum_{\mathclap{k'\ne k, k'=1}}^{K}\Tr\left(\mathbf{D}_{k'} \hat{\mathbf{H}}_k\right) +\sigma_k^2\Big) -\Tr\left(\mathbf{D}_k \hat{\mathbf{H}}_k\right)\Bigg)\nonumber \\
		 && + \sum_{m=1}^{M} \mu_m \left( \sum_{k=1}^{K} \Tr\left(\mathbf{Q}_m\mathbf{D}_k\right) -P_{\mathrm{max},m}\right),
		\ien
	    where $\lambda_k$, $\mu_k$, $\lambda_{k'}$ and $\mu_{k'}$ are non-negative Lagrange multipliers for user $k$ and all other users $k'$, respectively. The dual function is given by
	    \vspace{-0.1cm}
	    \ieb
	    &&g(\lambda_k, \mu_m) = \min_{\{\mathbf{D}_k\}} \mathcal{L}(\mathbf{D}_k, \lambda_k, \mu_k) \nonumber \\
	    &&= \kappa(\mathbf{z}) + \sum_{k=1}^{K} \lambda_k (2^{\tau_k}-1) \sigma_k^2  -\sum_{m=1}^{M} \mu_m P_{\mathrm{max},m}\nonumber \\
	    &&+ \min_{\{\mathbf{D}_k\}} \sum_{\mathclap{ k=1}}^{K} \Tr \left(\mathbf{Y}_k \mathbf{D}_k\right),
	    \label{dualf}
	   	    \ien	
		where 
	    \ieb
	    \mathbf{Y}_k = \hat{\mathbf{R}} + ~~~~\sum_{\mathclap{k'\ne k, k'=1}}^{K}\lambda_{k'}(2^{\tau_{k'}}-1) \hat{\mathbf{H}}_{k'} +\sum_{m=1}^{M}\mu_m\mathbf{Q}_m-\lambda_k \hat{\mathbf{H}}_k. 
	    \nonumber\\
	    \ien
	    	    
	    The minimum of \eqref{dualf} is $-\infty$ except for  $\mathbf{Y}_k\ge0, \forall k$. Thus, the Lagrange dual problem is
	   \ieb
	   &&\max_{\{\lambda_k,\mu_m\}} 	\kappa(\mathbf{z}) + \sum_{k=1}^{K} \lambda_k (2^{\tau_k}-1) \sigma_k^2  -\sum_{m=1}^{M} \mu_m P_{\mathrm{max},m} \nonumber \\
	   &\text{s.t.}~~~ &\mathbf{Y}_k\ge0,~~~~\forall k.
	   \label{dualp}
	   \ien
	   
	    Let $\lambda_k^*$ and $\mu_k^*, \forall k$ denote the optimal solutions for \eqref{dualp} and $\mathbf{D}_k^*$ be the optimal solution for \eqref{dualf}. Because $\mathbf{D}_k^* = \mathbf{d}_k^* (\mathbf{d}_k^*)^H $ and strong duality holds, the optimal digital precoders $\mathbf{d}_k^*$ can be obtained by
	    \vspace{-0.1cm} 
	   \ieb
	   \frac{\partial	\mathcal{L}(\mathbf{D}_k, \lambda_k^*, \mu_m^*)}{\partial \mathbf{d}_k} = \mathbf{Y}_k \mathbf{d}_k^* = \mathbf{0}.
	    \ien
	    
	    Since $ \mathbf{Y}_k$ is a block diagonal matrix, for each diagonal element, we have
	    \ieb
	    &&\Bigg(b_m\mathbf{R}_m^H\mathbf{R}_m+ ~~~\sum_{\mathclap{k'\ne k, k'=1}}^{K}\lambda_{k'}(2^{\tau _{k'}}-1) \mathbf{R}_{m}^H\mathbf{h}_{k',m}\mathbf{h}_{k',m}^H\mathbf{R}_{m}\nonumber \\
	    &&+\sum_{m'=1}^{M}\mu_{m'}^* \mathbf{Q}_{m,m'} -\lambda_k^* \mathbf{R}_{m}^H\mathbf{h}_{k,m}\mathbf{h}_{k,m}^H\mathbf{R}_{m}\Bigg)\mathbf{d}_{k,m}=\mathbf{0},
	    \label{asocond1}
	    \ien
	    hence, for a user $k$ to be served by BS $m$, the optimal digital precoder should satisfy
	    \ieb
	    \mathbf{d}^*_{k,m} =  c_{k,m}(\mathbf{B}_{k,m} )^{-1}\mathbf{R}_{m}^H\mathbf{h}_{k,m},
	    \label{asocond2}
	    \ien
	    where $c_{k,m} = \lambda^*_{k}\mathbf{h}_{k,m}^H\mathbf{R}_{m}\mathbf{d}^*_{k,m}$ and
	    \ieb
		\mathbf{B}_{k,m}&=& b_m\mathbf{R}_m^H\mathbf{R}_m + \sum_{\mathclap{k'\ne k, k'=1}}^{K}\lambda_{k'}(2^{\tau_{k'}}-1) \mathbf{R}_{m}^H\mathbf{h}_{k',m}\mathbf{h}_{k',m}^H\mathbf{R}_{m} \nonumber\\
		&&   
	    +\sum_{m'=1}^{M}\mu_{m'}^* \mathbf{Q}_{m,m'}. 
	    \ien
	    
	    Multiplying \eqref{asocond1} with $(\mathbf{d}^*_{k,m})^H$ from the left and plugging in \eqref{asocond2}, we have
	    \ieb
	    &&c_{k,m}^2\Bigg(\mathbf{h}_{k,m}^H\mathbf{R}_{m}(\mathbf{B}_{k,m})^{-1}\mathbf{R}_{m}^H\mathbf{h}_{k,m} \nonumber\\
	   && -\lambda_k^*\Big(\mathbf{h}_{k,m}^H\mathbf{R}_{m}(\mathbf{B}_{k,m})^{-1}\mathbf{R}_{m}^H\mathbf{h}_{k,m}\Big)^2\Bigg)= 0.
	    \label{multcond1}
	    \ien
	    
	     Hence, if a user $k$ should be served by BS $m$, the optimal multiplier should satisfy
	     \ieb
	     \lambda_k^* = \frac{1}{\mathbf{h}_{k,m}^H\mathbf{R}_{m}(\mathbf{B}_{k,m})^{-1}\mathbf{R}_{m}^H\mathbf{h}_{k,m}}, \label{multcond}
	     \ien
	     otherwise, the RF transmit power is set to 0 in order to satisfy \eqref{multcond1}.
	     Therefore, \eqref{multcond} is a necessary condition for user $k$ to be associated to BS $m$. Furthermore, according to the feasibility constraint $\mathbf{Y}_{k} \ge 0$, if we select $\mathbf{u}_{k,m}=\left(\mathbf{B}_{k,m}\right)^{-1} \mathbf{R}_{m}^H\mathbf{h}_{k,m}$, we have
	     \ieb
	     &&\mathbf{u}^H_{k,m}\mathbf{Y}_{k}\mathbf{u}_{k,m}=\mathbf{h}_{k,m}^H\mathbf{R}_{m}(\mathbf{B}_{k,m})^{-1}\mathbf{R}_{m}^H\mathbf{h}_{k,m}\nonumber \\
	     &&-\lambda_k^*\Big(\mathbf{h}_{k,m}^H\mathbf{R}_{m}(\mathbf{B}_{k,m})^{-1}\mathbf{R}_{m}^H\mathbf{h}_{k,m}\Big)^2 \ge 0.
	     \ien
	     
	     Thus, the Lagrange multiplier is found to satisfy
	     \vspace{-0.1cm}
	     \ieb
	     \lambda_k^* \le \frac{1}{\mathbf{h}_{k,m}^H\mathbf{R}_{m}(\mathbf{B}_{k,m})^{-1}\mathbf{R}_{m}^H\mathbf{h}_{k,m}}, \forall m,
	     \ien
	     where the equality holds for BSs with maximum $\mathbf{h}_{k,m}^H\mathbf{R}_{m}(\mathbf{B}_{k,m})^{-1}\mathbf{R}_{m}^H\mathbf{h}_{k,m}$.
	     Hence, the set of serving BSs of user $k$ is given by
	     \ieb
 	     \left\{m^*\Big|m^*=\argmax_{m} \mathbf{h}_{k,m}^H\mathbf{R}_{m}(\mathbf{B}_{k,m})^{-1}\mathbf{R}_{m}^H\mathbf{h}_{k,m}\right\}.
	     \label{associationR}
	     \ien
	     
	     The above expression gives the optimal user association principle in terms of the total transmit power. A user is served by multiple BSs if \eqref{associationR} has multiple elements, otherwise, the user is served by a single BS. 
	     According to \eqref{associationR}, the user association strategy is not simple as it is based on the analog precoders, the target spectral efficiency, the transmit power limit and the interference power. The serving BS of a user depends on the norm of the effective channel after analog precoding $\mathbf{R}_{m}^H\mathbf{h}_{k,m}$ which is weighted by $\mathbf{B}_{k,m}$ such that BSs with less interference power or higher transmit power limit are chosen. 
	  
		 \section{Sub-optimal cooperative hybrid beamforming algorithm}   
	     In this subsection, we propose a sub-optimal algorithm in terms of the silence strategy. The majority of the complexity is at the digital precoder optimization problem and the complexity scales with $\mathbf{z}_i$, therefore, we propose a sub-optimal algorithm that has a much lower complexity than $\mathcal{P}_1$ and it does not depend on $\mathbf{z}_i$. 
	     
	     Since $0\le P_m \le P_{\max,m}, \forall m$, the convex envelope of $P_m$ is given by
	     \vspace{-0.1cm}
	     \ieb
	     P_m^{\mathrm{sub}} = aP^{\mathrm{hw}}_m + \hat{\eta}_mP^{\mathrm{tx}}_{m},
	     \ien
	     where
	     \vspace{-0.1cm}
	     \ieb
	     \hat{\eta}_m = \frac{(1-a)P^{\mathrm{hw}}_m}{P_{\max,m}} + \frac{1}{\eta_m(1-\Delta_m)},
	     \ien
	     and the convex envelope of $P_m$ satisfies $ aP^{\mathrm{hw}}_m \le P_m^{\mathrm{sub}} \le \eta'_m P^{\mathrm{tx}}_{m} + P^{\mathrm{hw}}_m$. Next, we replace $ P_m$ with $ P_m^{\mathrm{sub}}$ in the objective function \eqref{objective},  and follow the same procedure as in Sec. \ref{secdp}. We then obtain the following convex optimization problem with a modified objective function:
	     \vspace{-0.1cm}	       
	       \ieb
	       &&\mathcal{P}_2 : \min_{\{\mathbf{D}_k\}} \sum_{k=1}^{K} \Tr\left( \hat{\mathbf{R}}(\hat{\eta}_m) \mathbf{D}_k\right) + a\sum_{m=1}^{M}b_mP^{\mathrm{hw}}_m 
	       \label{sub_obj}\\
	       &\text{s.t.}~~~&\Tr\left(\mathbf{D}_k \hat{\mathbf{H}}_k\right)\ge(2^{\tau_k}-1)\sum_{\mathclap{k'\ne k, k'=1}}^{K}\Tr\left(\mathbf{D}_{k'} \hat{\mathbf{H}}_k\right)\nonumber\\
	       && +(2^{\tau_k}-1)\sigma_k^2, ~~~~ \forall k,\label{lemma1_rate}\\
	       &&\sum_{k=1}^{K} \Tr\left(\mathbf{Q}_m\mathbf{D}_k\right) \le P_{\mathrm{max},m}, ~~~~~ \forall m,
	       \ien
	     where $\hat{\mathbf{R}}(\hat{\eta}_m)$ is obtained by replacing $\eta'_m$ in \eqref{rdy} with $\hat{\eta}_m $.
	     
	        \begin{algorithm}[t]
	     	\caption{Sub-optimal hybrid precoding for Cooperative MmWave Networks}
	     	\label{algo_sub}
	     	\begin{algorithmic}[1]
	     		\Require{$\mathbf{h}_{k,m}, \forall k,m$, $\tau_k,\forall k$, $P_{\mathrm{max},m} \forall m$, $\mathbf{r}_{k,m}$ or $\hat{\mathbf{r}}_{k,m}, \forall k,m$} according to Algorithm 1, $\epsilon$, stopping criterion $\epsilon_s$. 
	     		
	     		\State  Solve $\mathcal{P}_2$, obtain $\mathbf{D}_k^{(0)}$, $i=0$.
	     		\For{ $i= i+1$}
	     		\State Calculate the RF transmit power $P^{\mathrm{tx}, (i-1)}_{m} = \sum_{k=1}^{K}\Tr\left(\mathbf{Q}_m\mathbf{D}_k^{(i-1)}\right), \forall m$. 
	     		
	     		Define
	     		\ieb
	     		\hat{\eta}_m^{(i)} = \frac{(1-a)P^{\mathrm{hw}}_m}{ P^{\mathrm{tx}, (i-1)}_{m} + \epsilon} + \frac{1}{\eta_m(1-\Delta_m)},
	     		\ien
	     		where $\epsilon \ll P^{\mathrm{tx}, (i-1)}_{m}$ and the modified total power for BS $m$ is given by $ P_m^{\mathrm{sub}} = aP^{\mathrm{hw}}_m + \hat{\eta}_m^{(i)}P^{\mathrm{tx,(i)}}_{m}$.
	     		\State Solve $\mathcal{P}_2$ with modified objective function, 
	     		\ieb
	     		\sum_{k=1}^{K} \Tr\left( \hat{\mathbf{R}}\left(\hat{\eta}_m^{(i)}\right) \mathbf{D}_k\right) + a\sum_{m=1}^{M}b_mP^{\mathrm{hw}}_m	,
	     		\ien
	     		
	     		obtain $\mathbf{D}_k^{(i)}$.
	     		\State If  $\sum_{m=1}^{M}|P^{\mathrm{tx}, (i)}_{m} - P^{\mathrm{tx}, (i-1)}_{m}| < \epsilon_s$, go to 7.
	     		
	     		\EndFor
	     		
	     		\State Obtain the sub-optimal digital precoders $\mathbf{D}_k^* = \mathbf{D}_k^{(i)}$ in terms of the BS silence strategy.
	     		\State Sum RF transmit power  $P_{\mathrm{tx, sub}}^*=\sum_{m=1}^{M}\sum_{k=1}^{K} \Tr\left( 	\mathbf{Q}_m \mathbf{D}_k^*\right)$ ,
	     		\State Sum power $ P^*_{\mathrm{sub}} =\sum_{k=1}^{K} \Tr\left( 	\hat{\mathbf{R}}\left(\hat{\eta}_m^{(i)}\right) \mathbf{D}_k^*\right)+ a\sum_{m=1}^{M}b_mP^{\mathrm{hw}}_m.$ 
	     	\end{algorithmic}
	     \end{algorithm}
	     
	     In order to drive BSs into the silent mode, we propose the iterative sub-optimal mmWave hybrid beamforming algorithm (Algorithm 2) that modifies the objective function at each iteration such that BSs with small RF transmit power are driven into the silent mode. We first obtain $P^{\mathrm{tx}, (0)}_{m}$ by solving $\mathcal{P}_2$ using $\hat{\eta}_m$. Then, for each iteration, if $P^{\mathrm{tx}, (i-1)}_{m}$ is small, $\hat{\eta}_m^{(i)}$ is large, which makes the sum power $ \sum_{k=1}^{K} \Tr\left( \hat{\mathbf{R}}\left(\hat{\eta}_m^{(i)}\right) \mathbf{D}_k\right) + a\sum_{m=1}^{M}b_mP^{\mathrm{hw}}_m$ large. Therefore, in the next optimization problem, BS $m$ will be driven to the silent mode. For BS $m$, the sub-optimal RF transmit power is given by $P_{m,\mathrm{tx}}^*=\sum_{k=1}^{K}\Tr\left( 	\mathbf{Q}_m \mathbf{D}_k^*\right)$. If $P_{m,\mathrm{tx}}^*>> \epsilon$, the total power of BS $m$ is given by $P_{m,\mathrm{sub}}^* \approx aP^{\mathrm{hw}}_m + \left(\frac{(1-a)P^{\mathrm{hw}}_m}{ P_{m,\mathrm{tx}}^* + \epsilon} + \frac{1}{\eta_m(1-\Delta_m)}\right) P_{m,\mathrm{tx}}^* \approx aP^{\mathrm{hw}}_m + \eta_m'P_{m,\mathrm{tx}}^*$ and the BS is in active mode. If $P_{m,\mathrm{tx}}^* = 0$, the BS is in silent mode, and the total power is given by  $P_{m,\mathrm{sub}}^* = aP^{\mathrm{hw}}_m$.
	     Also, Algorithm 2 is sub-optimal to Algorithm 1, in terms of the BS silence strategy, since Algorithm 1 obtains the optimal BS silence strategy through exhaustive search. Moreover, it is worth mentioning that Algorithm 2 always converges. The proof can be found in \cite[Appendix B]{subop_proof}. 
	     
	     Finally, we want to mention that the complexity of Algorithm 2 is given by replacing $(2^M-1)$ in the complexity of Algorithm 1 with the number of iterations $i$. In our simulations, the average number of iterations of FHP and FDP obtained is approximately 1 for different target spectral efficiency. For PHP, the average number of iterations is in the range (1,2) with the average number of iterations increasing with the target spectral efficiency. Therefore, the complexity of Algorithm 2 is low.

	    \section{Cooperative Hybrid Beamforming for OFDM Systems}
	    OFDM is chosen to be the main multi-carrier waveform in 5G cellular systems \cite{3GPP_OFDM}, because it is resilient to frequency selective fading while providing high spectral efficiency.
	    In this subsection, we extend the cooperative hybrid precoding to the case of frequency selective channels and show the analog and digital precoders solutions for OFDM. 
	    
	    We consider a hybrid precoding system that employs OFDM as the multi-carrier transmission waveform with $N_\mathrm{s}$ sub-carriers. At BS $m$, the data symbol $x_{k,m}[n_\mathrm{s}]$ for user $k$ at sub-carrier $n_\mathrm{s}$ is first precoded by the digital precoder $\mathbf{d}_{k,m}[n_\mathrm{s}], 0\le n_\mathrm{s}< N_\mathrm{s}$. Next, a symbol block of symbols modulated on different sub-carriers is transformed to the time-domain symbols via an Inverse Fast 
	    Fourier Transform (IFFT) and the cyclic prefix is added to the beginning of the symbols. Then, the OFDM symbol is precoded by the analog precoder $R_m$ which is the same for all sub-carriers. For FDP, the analog precoder is not needed. 
	    
	    To incorporate the multi-carrier waveform, we adopt a frequency domain channel model \cite{Sohrabi_OFDM, Tsinos_energyefficiency} for sub-carrier $n_\mathrm{s}$	
	    \ieb
	    \mathbf{h}_{k,m}[n_s] = \sqrt{\frac{\rho_{k,m}N_m}{N_{\mathrm{cl}}N_{\mathrm{ray}}}}
	    \sum_{i=1}^{N_{\mathrm{cl}}}\sum_{l=1}^{N_{\mathrm{ray}}}\alpha_{i,l}
	    \mathbf{a}^t_m(\theta^t_{i,l}) e^{-j\frac{2\pi n_\mathrm{s}}{N_\mathrm{s}}i},
	    \label{fschannela}	
	    \ien
	    where the definition of parameters are according to \eqref{nbchannel}.

	     Similar as the single-carrier case, we assume that the BS can jointly serve a user on certain sub-carriers or leave some sub-carriers empty to reduce the interference.
	    The composite signal received by user $k$ on subcarrier $n_\mathrm{s}$ can be written as
	    \ieb
	    y_k[n_s] &=& \sum_{m=1}^M \mathbf{h}^H_{k,m}[n_\mathrm{s}] \mathbf{R}_{m}\mathbf{d}_{k,m}[n_\mathrm{s}]x_{k,m}[n_\mathrm{s}]+\sum_{m=1}^{M}\sum_{k'\ne k,k'=1}^K \nonumber\\
	    &&\mathbf{h}^H_{k,m}[n_\mathrm{s}]\mathbf{R}_{m}\mathbf{d}_{k',m}[n_\mathrm{s}]x_{k',m} [n_\mathrm{s}]
	    + n_k[n_\mathrm{s}].
	    \ien

	    Here, $\mathbf{R}_{m}$ is the analog precoder for BS $m$ and for all subcarriers, $\mathbf{d}_{k,m}[n_\mathrm{s}]$ and $x_{k,m}[n_\mathrm{s}]$ are the digital precoder and the transmit symbol of subcarrier $n_\mathrm{s}$, respectively. Assuming Gaussian signaling, the sum spectral efficiency for user $k$ at subcarrier $n_\mathrm{s}$ is given by
	    \ieb
	    \Gamma_k[n_\mathrm{s}]=\log_2 \left(1+\frac{\sum_{m=1}^{M}|\mathbf{h}_{k,m}^H[n_\mathrm{s}]\mathbf{w}_{k,m}[n_\mathrm{s}]|^2}{I_k[n_\mathrm{s}]+\sigma_k^2}\right).
	    \ien
	    where 
	    $I_k[n_\mathrm{s}]=\sum_{m=1}^{M}\sum_{k'\ne k,k'=1}^{K}\mathbf{w}_{k',m}^H[n_\mathrm{s}]\mathbf{h}_{k,m}[n_\mathrm{s}]\mathbf{h}_{k,m}^H[n_\mathrm{s}]\\ \times\mathbf{w}_{k',m}[n_\mathrm{s}]$ and  
	    $\mathbf{w}_{k,m}[n_\mathrm{s}]= \mathbf{R}_{m}\mathbf{d}_{k,m}[n_s].$

	    The cooperative hybrid beamforming problem for minimizing the total power consumption for the frequency selective channel case can be cast in the same form of $\mathcal{P}$, which is given by
	       		\ieb
	   \mathcal{P}_3:  && \min_{\{\mathbf{R}_{m},\mathbf{d}_{k,m}[n_\mathrm{s}]\}}\sum_{m=1}^{M} b_mP_m[n_\mathrm{s}] \\
	    &&\mathrm{s.t.}~~~ \Gamma_k[n_\mathrm{s}]\ge \tau_k[n_\mathrm{s}], ~~~~~ \forall k \label{fs_c1}\\
	    &&\sum_{k=1}^{K}||\mathbf{R}_{m}\mathbf{d}_{k,m}[n_s]||^2 \le P_{\max,m}[n_s],~~~~~ \forall m \label{fs_c2}\\
	    &&\left|\mathbf{R}_{m}^{(i,j)}\right|=\frac{1}{\sqrt{N_mL_m}}, ~~~~~ \forall m,i,j,~~~~~\mathrm{for}~\mathrm{FHP}, \\
	    &&\mathbf{R}_m = \mathrm{diag}[\hat{\mathbf{r}}_{1,m}, \cdots, \hat{\mathbf{r}}_{L_m,m}],\nonumber \\
	    &&\left|\hat{\mathbf{r}}_{i,m}^{(j)}\right|= 1/\sqrt{N_m},  ~~~ \forall m,j, ~~~\mathrm{for}~\mathrm{PHP}. 
	    \ien
	    
	    Here, $P_m[n_\mathrm{s}]$ is according to \eqref{powermodel} for sub-carrier $n_s$, \eqref{fs_c1} is the per sub-carrier rate constraint with minimum spectrum efficiency $\tau_k[n_\mathrm{s}]$ and \eqref{fs_c2} is the per-subcarrier maximum power constraint. 
	    
	    It is worthwhile to mention that we assume a per sub-carrier power constraint, which is a hard constraint and the results may serve as a lower bound for problems with other more relaxed rate constraints. It implies that we can adjust the peak power individually for each sub-carrier and make sure the PAPR is acceptable. Without
	    the per sub-carrier power constraint, we can use the power amplifier efficiency $\eta_m$
	    to evaluate the effect of PAPR. For example, for high PAPR, we can set the power amplifier efficiency $\eta_m$ to a small number, which increases the power consumption of each transmitting BS.
	    Also, $\mathcal{P}_3$ indicates that the digital precoders are optimized per sub-carrier while the analog precoder is fixed for all sub-carriers. Therefore, following the similar digital and analog precoder decoupling steps as in the single-carrier case, we start by finding the analog precoder for all sub-carriers. Then, conditioned on the analog precoder, we solve the digital precoder problem per sub-carrier.
	    
	    Following the analog precoding problem in the single-carrier system of maximizing the array gain as in \eqref{EGT_problem} and \eqref{EGT_probcon}, we aim to find the analog precoder in the multi-carrier narrow-band system which maximizes the sum of array gain over all sub-subcarriers. Hence, for FHP, the analog precoding problem in the OFDM system can be written as
	    	\ieb
	    &&\max_{\mathbf{r}_{k,m}}\sum_{n_s=0}^{N_\mathrm{s}-1}\left|\mathbf{h}^H_{k,m}[n_s]\mathbf{r}_{k,m}\right|^2 ~~~~~ \\
	    &&\mathrm{s.t.}\left|\mathbf{r}_{k,m}^{(i)}\right|=\frac{1}{\sqrt{L_mN_m}}, ~~~~~ 0< i\le N_m,
	    \ien
	    where $\mathbf{r}_{k,m}$ is the $k$-th column of $\mathbf{R}_m$.
	    
	    According to the Cauchy–Schwarz inequality, we have
	    \ieb
	   \sum_{n_s=0}^{N_s-1}\left|\frac{1}{\sqrt{N_\mathrm{s}}} \right|^2\sum_{n_s=0}^{N_s-1}\left|\mathbf{h}^H_{k,m}[n_\mathrm{s}]\mathbf{r}_{k,m}\right|^2 &\ge& \frac{1}{N_s} \nonumber\\
	   \times\left| \sum_{n_s=0}^{N_s-1} \mathbf{h}^H_{k,m}[n_s]\mathbf{r}_{k,m} \right|^2. 
	    \ien
	    Defining $\bar{\mathbf{h}}_{k,m}^H= \sum_{n_s=0}^{N_s-1} \mathbf{h}^H_{k,m}[n_s]$ and considering maximizing the above lower bound, the optimization problem of finding $\mathbf{R}_m$ can be written as
	    \ieb
	     &&
	     \label{fs_a1}	\max_{\{\mathbf{r}_{k,m}\}}\left|\bar{\mathbf{h}}^H_{k,m}\mathbf{r}_{k,m}\right|^2 ~~~~~ \\
	     \label{fs_a2}
	    &&\mathrm{s.t.}\left|\mathbf{r}_{k,m}^{(i)}\right|=\frac{1}{\sqrt{L_mN_m}}, ~~~~~ 0< i\le N_m.
	    \ien
	    Since \eqref{fs_a1} and  \eqref{fs_a2} are in the same form as \eqref{EGT_problem} and \eqref{EGT_probcon}, the analog precoder has a closed form solution given by
	    \vspace{-0.1cm}  
	    \ieb
	    &&\mathbf{r}_{k,m}^{(i)}=\frac{1}{\sqrt{L_mN_m}}e^{j\left(\xi+ \angle\bar{\mathbf{h}}_{k,m}^{(i)}\right)}.
	    \ien
	    
	    Following the same procedure as above, the analog precoder for PHP is given by
	    \ieb
	    &&\hat {\mathbf{r}}_{k,m}^{(i)}=\frac{1}{\sqrt{N_m}}e^{j\left(\xi+ \angle \bar{\mathbf{h}}_{k,m}^{((k-1)N_m/L_m+i)}\right)},  0<i \le N_m/L_m,\nonumber \\
	    \ien
	     where $\hat{\mathbf{h}}^H_{k,m}= \sum_{n_s=0}^{N_s-1} \mathbf{h}^H_{k,m}[n_s] \mathbf{G}_{k,m}$ and $\mathbf{G}_{k,m}$ is given by \eqref{RF_G}.
	    
	    Once the analog precoder is determined, it is straightforward to see that $\mathcal{P}_3$ can be cast in the same form as $\mathcal{P}_1$ and be solved by Algorithm 1. It is
	    noteworthy that the digital precoder optimization problem in the case of OFDM is decoupled into $N_s$ independent problems, one for each sub-carrier. After Algorithm 1, the precoders will jointly assign sub-carriers for each BS-user pair. Some sub-carriers might be silenced due to interference suppression or bad channel conditions and some sub-carriers might be shared among BSs to achieve joint transmissions.

		\section{Simulation Results}

\begin{figure}
		 	\centering
		 	\begin{subfigure}[b]{0.23\textwidth}
		 		\centering
		 		\includegraphics[width=\linewidth]{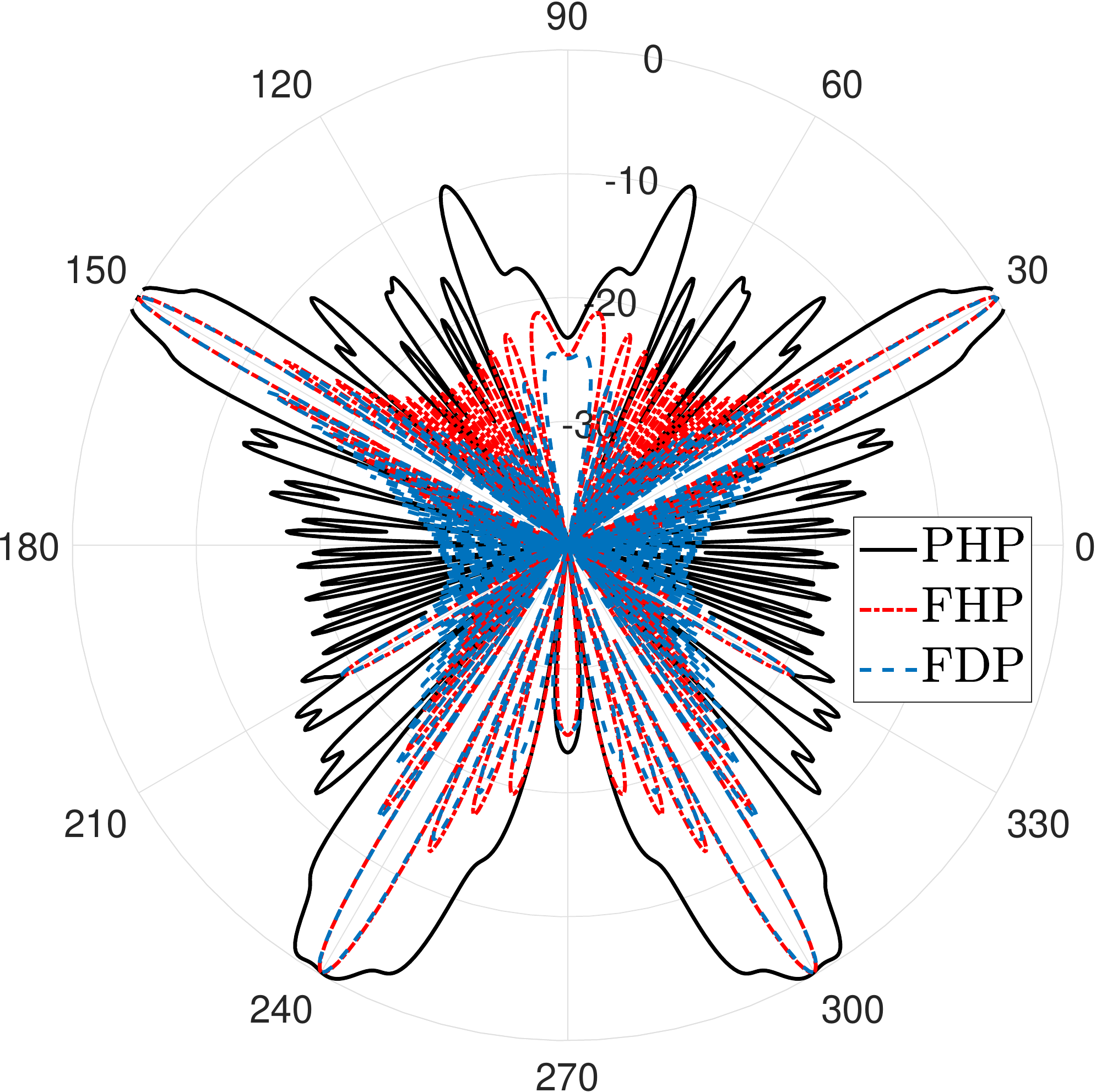}
		 		\caption{BS1, $N_1 = 64$.}
		 		\label{pattern_64a}
		 	\end{subfigure}
		 	\hfill
		 	\begin{subfigure}[b]{0.23\textwidth}
		 		\centering
		 		\includegraphics[width=\linewidth]{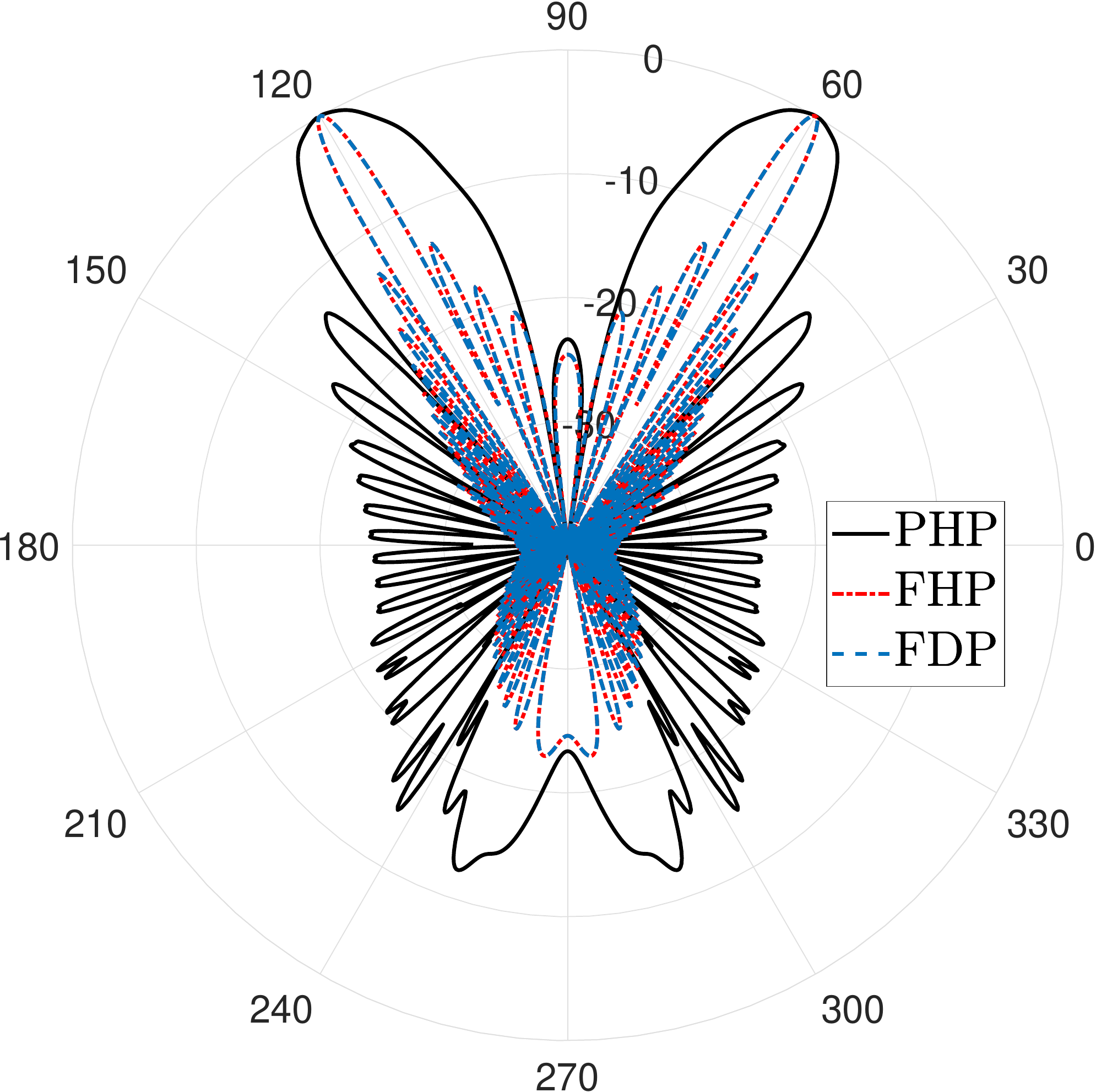}
		 		\caption{BS2, $N_2 = 64$.}
		 		\label{pattern_64b}
		 	\end{subfigure}
		 	\hfill
		 	\begin{subfigure}[b]{0.23\textwidth}
		 		\centering
		 		\includegraphics[width=\linewidth]{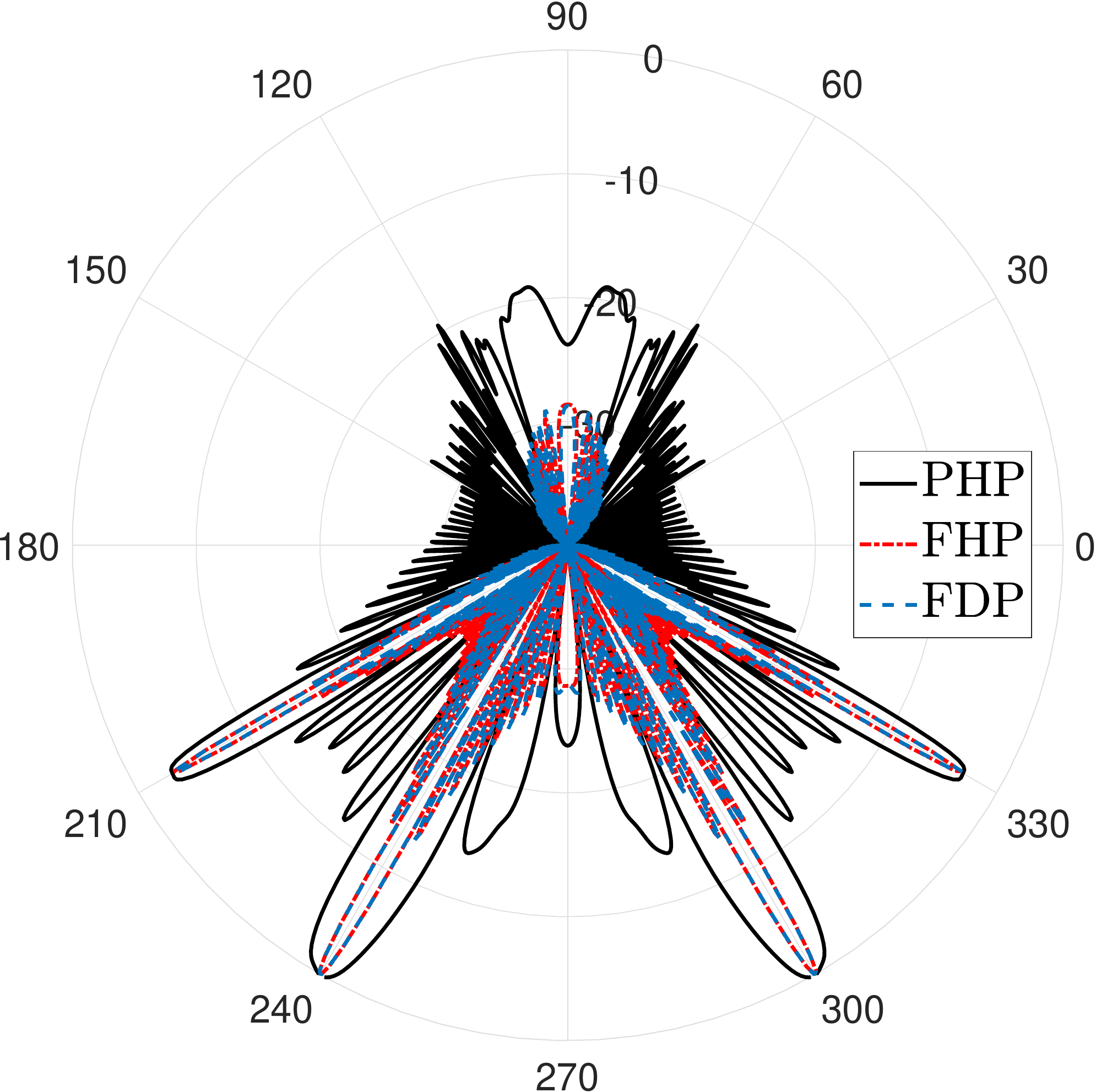}
		 		\caption{BS1, $N_1 = 128$.}
		 		\label{beam_pattern3}
		 	\end{subfigure}
		 	\hfill
		 	\begin{subfigure}[b]{0.23\textwidth}
		 		\centering
		 		\includegraphics[width=\linewidth]{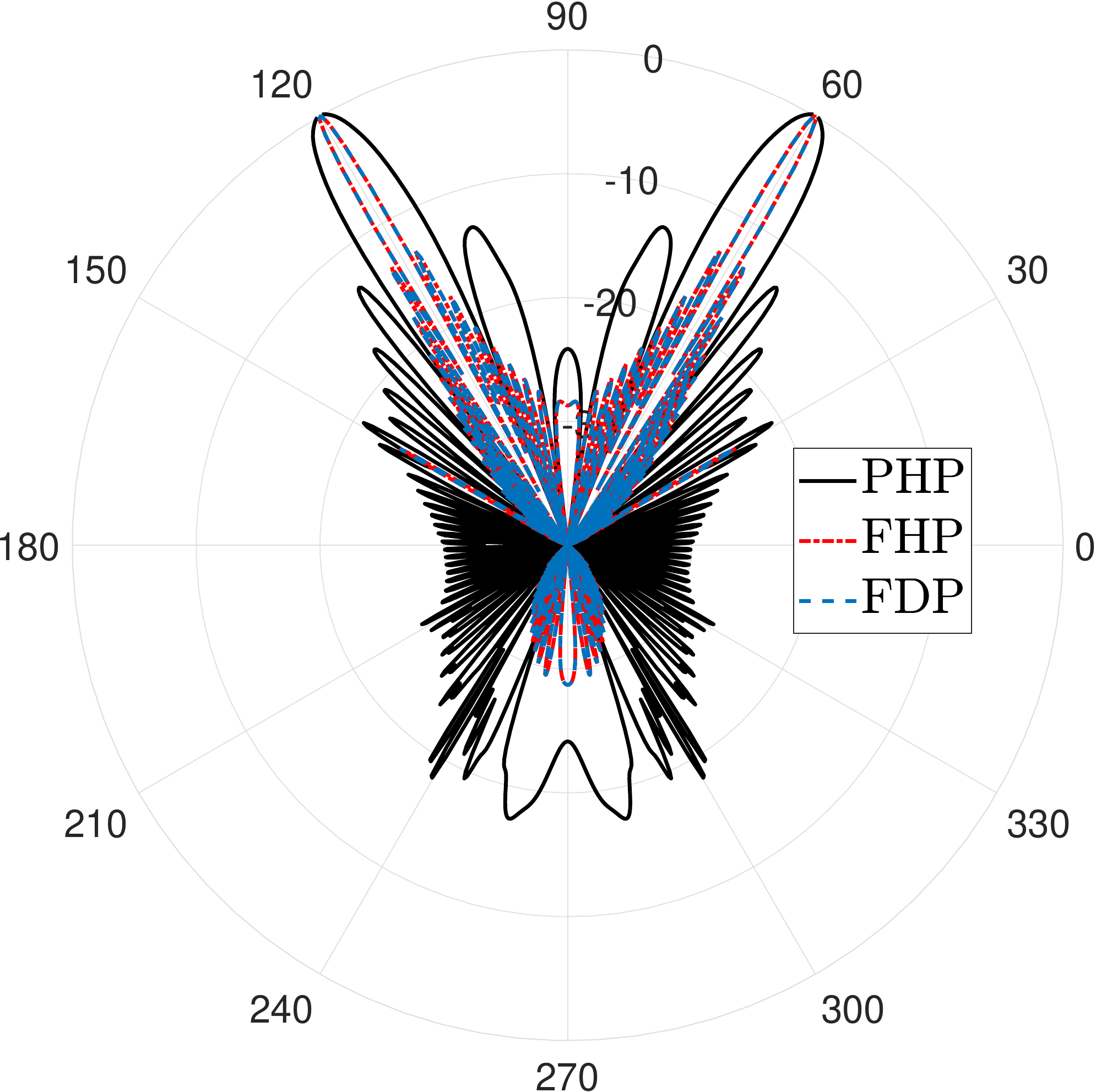}
		 		\vspace{-0.2cm}
		 		\caption{BS2, $N_2 = 128$.}
		 		\label{beam_pattern4}
		 	\end{subfigure}
		 	\caption{Beam patterns in a network with $M=2$ BSs and $K=4$ users. All BSs are set to be active, the number of RF chains is $L_m=4$ per BS and the target spectral efficiency is $\tau_k =4$ bit/s/Hz per user. The AOD is assumed to be at $-60^{\circ}, -30^{\circ}, 30^{\circ}, 60^{\circ}$. The beam gains are shown in dB scale and are normalized with respect to the largest beam gain in the FDP. (a) and (b): Beam patterns at BS1 and BS2 with 64 antennas, respectively. (c) and (d): Beam patterns at BS1 and BS2 with 128 antennas, respectively.}
		 	\label{beam_pattern}
		 \end{figure}

\begin{figure}[t]
	    	\centering
	    	\includegraphics[width=\linewidth]{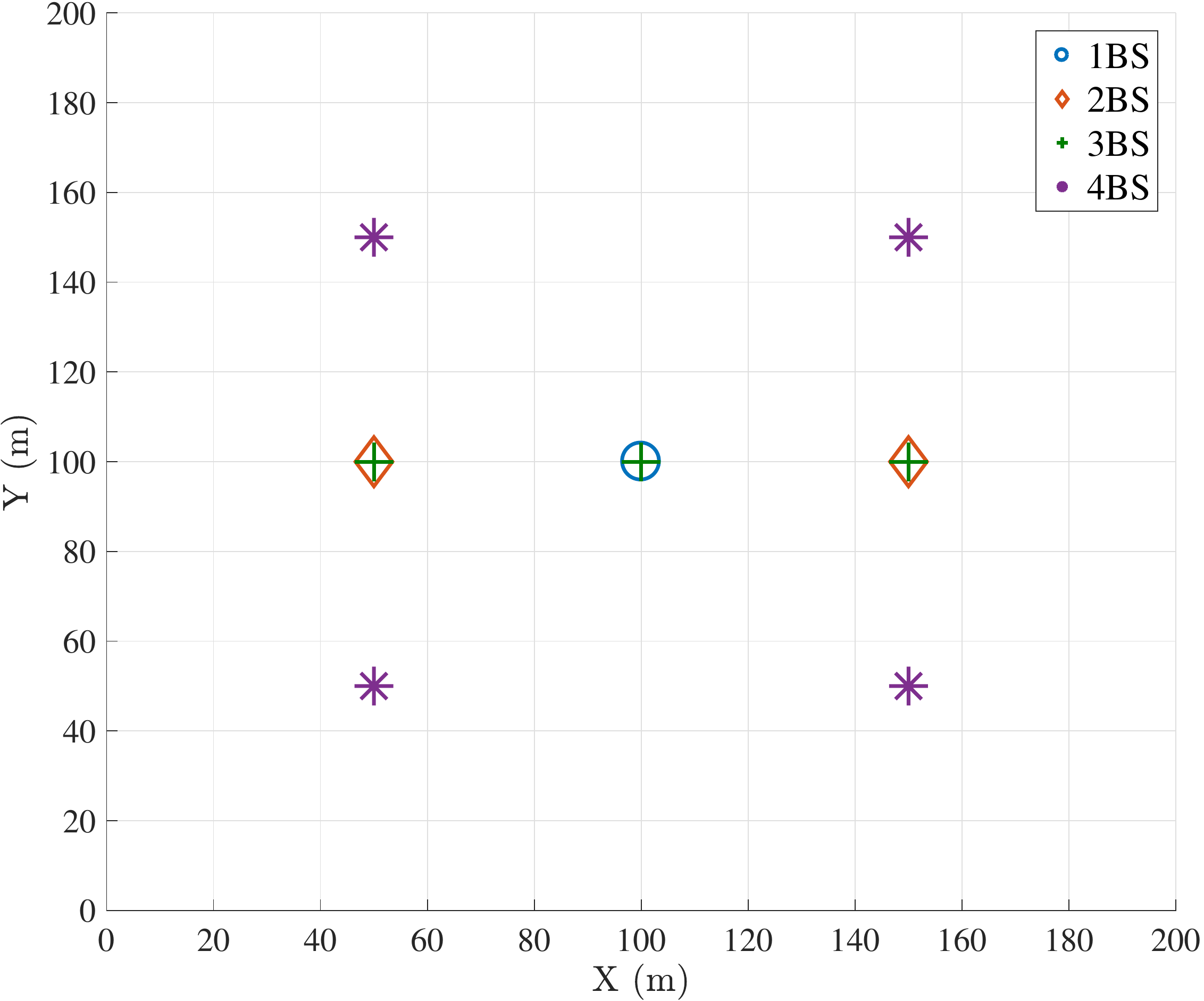}
	    	\vspace{-0.2cm}
	    	\caption{The locations of the BSs for $M={1,2,3,4}$. For $M=5$, the locations of the BSs is a combination of that of 1 BS and 4 BSs.}
	    	\label{bslocation}
	    \end{figure}

		In this section, we present the simulation results for the FHP, the PHP and the FDP.  we first compare the beam patterns of the FHP, the PHP and the FDP (Fig.~\ref{beam_pattern}). Then, we 
		show the RF power consumption of the FHP, the PHP, the FDP and their corresponding sub-optimal cases (Fig.~\ref{partial_txpower}). Also, the effect of hardware power consumption on the sum power consumption is shown for the FHP and the PHP (Fig.~\ref{an_active}). Finally, we show how the BS cooperation can  reduce the sum RF transmit power, the sum power of the BSs, as well as the infeasible solutions (Figs. \ref{power_hl}-\ref{powerCDF_bs}), and increase energy efficiency of OFDM systems for the FHP and the PHP (Fig.~\ref{EE_OFDM}). We choose hardware power consumption values given in Table~\ref{simulation_parameters} for reference and set the weighting parameter $b_m=1, \forall m,$ and $\beta = 0.01$ unless otherwise specified. Such parameter settings are in harmony with, e.g., \cite{Abbas_HBADCComparison, Roi_PhaseShifterorSwitch}, and have been selected based on our discussion with Ericsson, so that we provide fair comparisons for different architectures. The simulation results are averaged over $10^5$ channel realizations, and in each realization users are randomly dropped in an area of 200 m $\times$ 200 m. The locations of the BSs vary according to the number of cooperative BSs. We consider fixed locations for BSs which are depicted in Fig.~\ref{bslocation}.

			\begin{table}[t]
			\centering
			\caption {Simulation parameters.}
			\label{simulation_parameters} 
			\begin{tabular}{|  p{0.7cm} | p{0.7cm} | p{0.7cm} | p{0.7cm} | p{0.5 cm} |  p{0.5 cm} |}
				\hline
				$P_{\mathrm{PS}}$ [mW]& $P_{\mathrm{DAC}}$ [mW]   & $P_{\mathrm{RF}}$ [mW]&   $P_{\text{max},m}$ [dBm] & $\eta_m$ & $\Delta_m$ \\
				\hline
				40 & 200 & 40 &55 & 0.3& 0.15 \\
				\hline
				 $N_{\textrm{cl}}$  & $N_{\textrm{ray}}$ & $\bar{n}_{\textrm{LOS}}$ & $\bar{n}_{\textrm{NLOS}}$ & $\sigma_{\textrm{LOS}}$ [dB] &  $\sigma_{\textrm{NLOS}}$ [dB]\\
				\hline
				 2 & 20 & 2.1 & 3.4 & 3.6 & 9.7 \\	
				\hline
			\end{tabular}
			\vspace{-4mm}
		\end{table}

		\subsection{On Beam Pattern}
	     \label{simu_beam}

		In Fig.~\ref{beam_pattern}, we compare the beam patterns of the FDP, the FHP, and the PHP for $M=2$ BSs, $K=4$ users and $L_m=4$ RF chains per BS. All BSs are set to be active and the target spectral efficiency is $\tau_k =4$ bit/s/Hz for all users. The shown beam patterns of different array sizes are based on different channel realizations and considering only the azimuth domain for simplicity. However, if the elevation domain is considered, more energy focusing beams can be formed and the total power consumption is expected to reduce.

		 In Figs.~\ref{beam_pattern}(a) and \ref{beam_pattern}(b), we can see how the two BSs jointly serve four users with main lobes pointed at AODs. In Figs.~\ref{beam_pattern}(a), the power of the beam pointed at $-30^{\circ}$ is -30 dB smaller than the peak beam power. This means that the user at $-30^{\circ}$ has a good channel condition and it should be LOS transmission, since the target spectral efficiency can be met without much beamforming power. Similar results can be observed in Fig~\ref{beam_pattern} (d). By comparing the beam patterns among different architectures, it confirms that our proposed FHP gives close performance to the FDP in terms of the main lobe angles and maximum gains. Also, we notice that the PHP generates wider beams and thus may cause more interference to other directions. Figures \ref{beam_pattern}(c) and \ref{beam_pattern}(d) show that, by increasing the number of antennas, the beam main lobes become narrower and more energy focusing, especially for the PHP. Hence, if a large antenna array is available, it is possible to achieve energy-focusing narrow beams with PHP, while keeping the number of PSs and complexity low.  	

		\subsection{On Power Consumption}
		\label{simu_power}
		In this subsection, we assess the power consumption of different architectures and show the effect of the number of antennas and PSs on the power consumption.
		\begin{figure}[t]
			\centering
			\includegraphics[width=\linewidth]{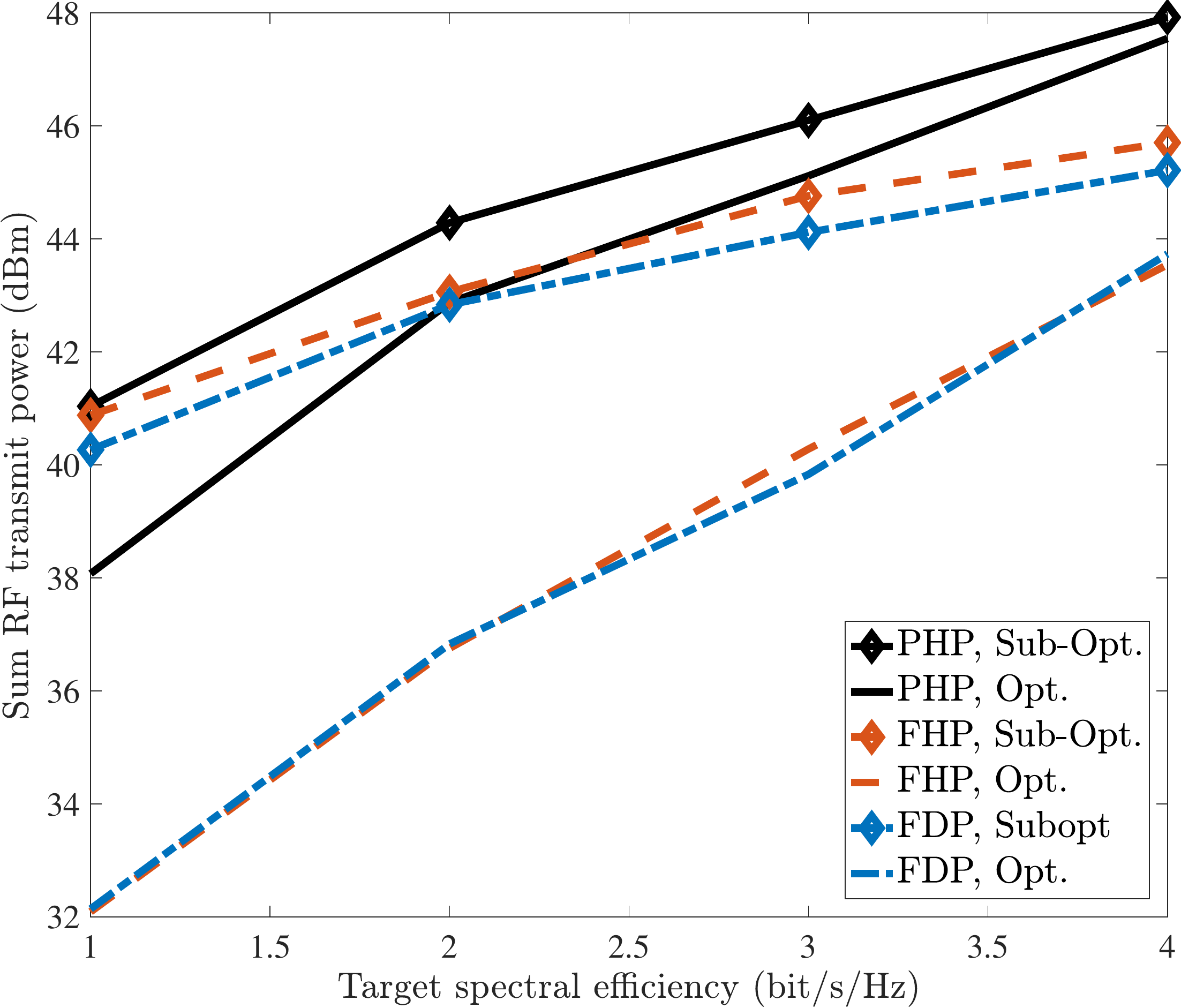}
			\caption{Sum RF transmit power $P_{\mathrm{tx}}^*$ with the optimal BS silent mode and the sub-optimal BS silent mode. 
			`Opt.' denotes the cases with the optimal BS silent mode which is obtained from Algorithm 1. `Sub-Opt.' denotes the cases with sub-optimal BS silent mode which is obtained from Algorithm 2.	
				The network parameters are $M=2$ BSs, $K=4$ users, $N_m=64$ antennas and $L_m=4$ RF chains per BS.}
			\label{partial_txpower}
		\end{figure}	
		
		Setting $M=2$ BSs, $K=4$ users, $N_m=64$ antennas, $L_m=4$ RF chains per BS, Fig.~\ref{partial_txpower} shows the sum RF transmit power consumption $P_{\mathrm{tx}}^*$ versus target spectral efficiency. It confirms that, for a
		given RF transmit power, the FHP scheme can achieve close spectral efficiency compared to the FDP and this also applies to the  sub-optimal cases. \footnote{The fluctuations around 3 and 4 bit/s/Hz are due to some remaining statistical averaging errors and randomness of the channel realizations, caused by the large amount of involved random realizations that makes it difficult to obtain fully smooth curves.} However, the PHP needs to transmit with more power than the FDP or the FHP to achieve a given target spectral efficiency. Intuitively, this is because the reduced number of RF chains and PSs leads to less control over the precoders. In order to satisfy a high target spectral efficiency, it becomes difficult for the PHP to form as energy-focusing beams as the FDP or the FHP, thus the PHP needs to increase the RF transmit power. For all architectures, the gap between the sub-optimal case and the optimal case decreases with the target spectral efficiency. This is because more BSs need to be active to achieve a higher target spectral efficiency. Both the optimal and sub-optimal case are driven towards the all active case and the difference in terms of power consumption becomes small. For PHP, since the BS activation probability of both the optimal case and the sub-optimal case is high and similar, the gap between the sub-optimal case and the optimal case is smaller than that of the FDP and FHP. Also, our simulations show that the same results are observed for the sum total power.
		
		\begin{figure}[t]
			\centering
			\includegraphics[width=\linewidth]{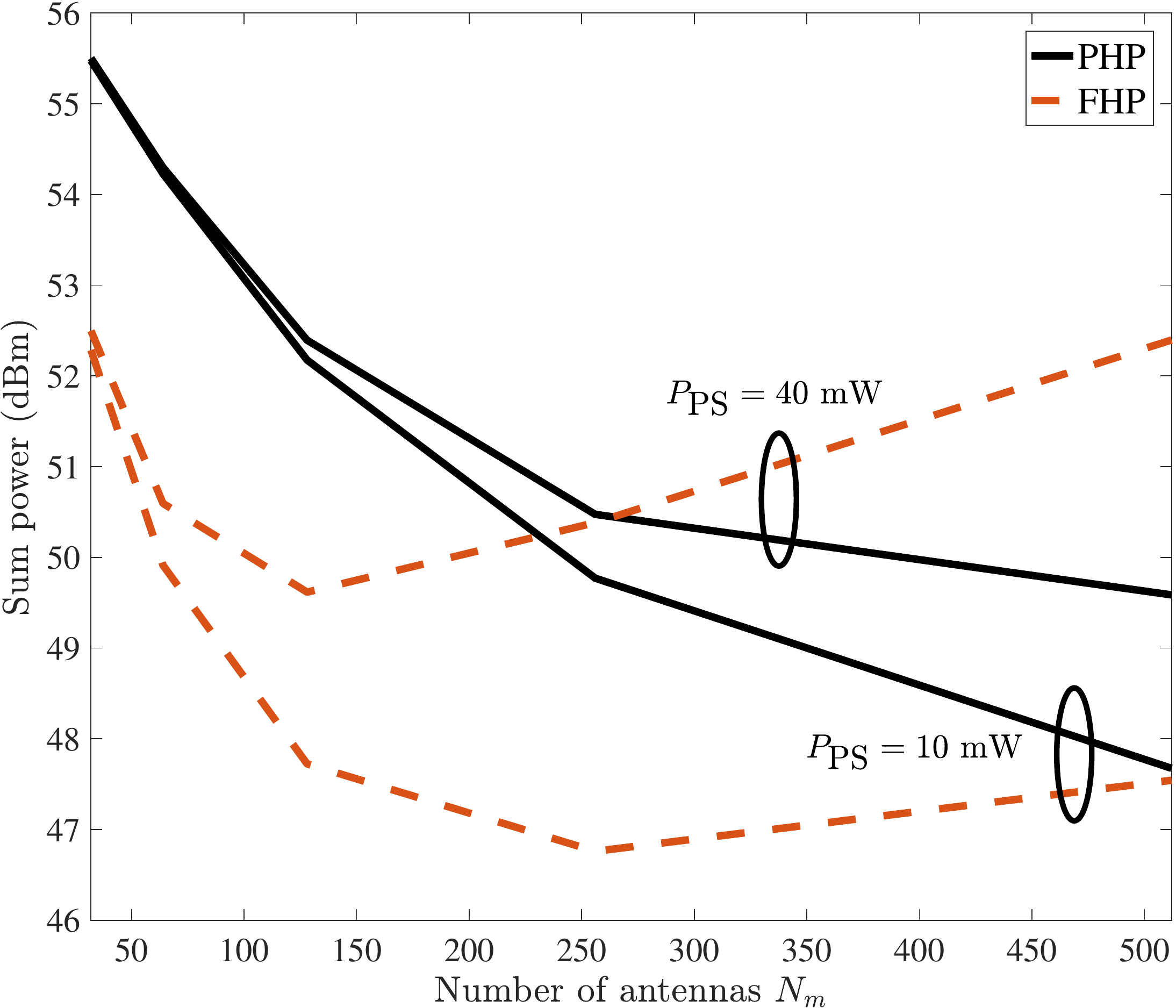}
			\caption{ Sum power consumption $P^*$ versus the number of antennas 
				with the optimal BS silence pattern. The network parameters are $M=2$ BSs, $K=4$ users, $L_m=4$ RF chains per BS and the target spectral efficiency $\tau_k = 4$ bit/s/Hz per user. For each architecture, the results are simulated based on two values of PS power consumption $P_{\mathrm{PS}}= 10$ mW and $P_{\mathrm{PS}}= 40$ mW.}
			\label{an_active}
		\end{figure}

		To examine the effect of antennas and PSs on the power consumption, we compare the sum power consumption
		for two PS power values $P_{\mathrm{PS}}= 10$ mW and $P_{\mathrm{PS}}= 40$ mW in Fig.~\ref{an_active}. Here, the results are presented for the cases with the number of BSs $M=2$, the number of users $K=4$, $L_m=4$ RF chains per BS and the target spectral efficiency $\tau_k = 4$ bit/s/Hz per user, the number of antennas is chosen such that $N_m/L_m$ is an integer. 
		We exclude the case with FDP because the size of the precoding matrix grows exponentially with the number of antennas, causing memory overflow and resulting in optimization problems with prohibitive number of parameters.

		Figure \ref{an_active} shows that, for small array sizes, the sum power consumption of the FHP decreases with the number of antennas as a result of increased beamforming gains and, thus, reduced RF transmit power. However, the sum power consumption starts to increase for larger number of antennas and the rate of increase is larger for a larger $P_{\mathrm{PS}}$. This is intuitively so because the power consumption is dominated by the increased hardware power when the number of antennas is large. 
		Also, Fig.~\ref{an_active} shows that the sum power consumption of the PHP keeps decreasing with the number of antennas since the number of PSs needed in the PHP increases less quickly than that of the FHP and the power consumption is dominated by the decreasing RF transmit power. Therefore, depending on how fast the hardware power increases with the number of antennas, there exists a cross-over point where the PHP outperforms the FHP, in terms of the sum power consumption for a given target spectral efficiency.
		The result indicates that, in order for the PHP to consume less power than the FHP, the RF transmit power gap between the PHP and the FHP needs to be reduced, which can be achieved by increasing the number of antennas or using overlap in terms of RF chains between subarrays such that a larger beamforming gain is achieved.
	
	\subsection{On the Value of Cooperation}
	\label{simu_coop}
	    In this subsection, we show simulation results for different number of BSs and analyze the value of cooperation, in terms of the power consumption, infeasibility probability, as well as the BS cooperation probability.
    
    \begin{figure}[t]
    	\centering
    	\includegraphics[width=\linewidth]{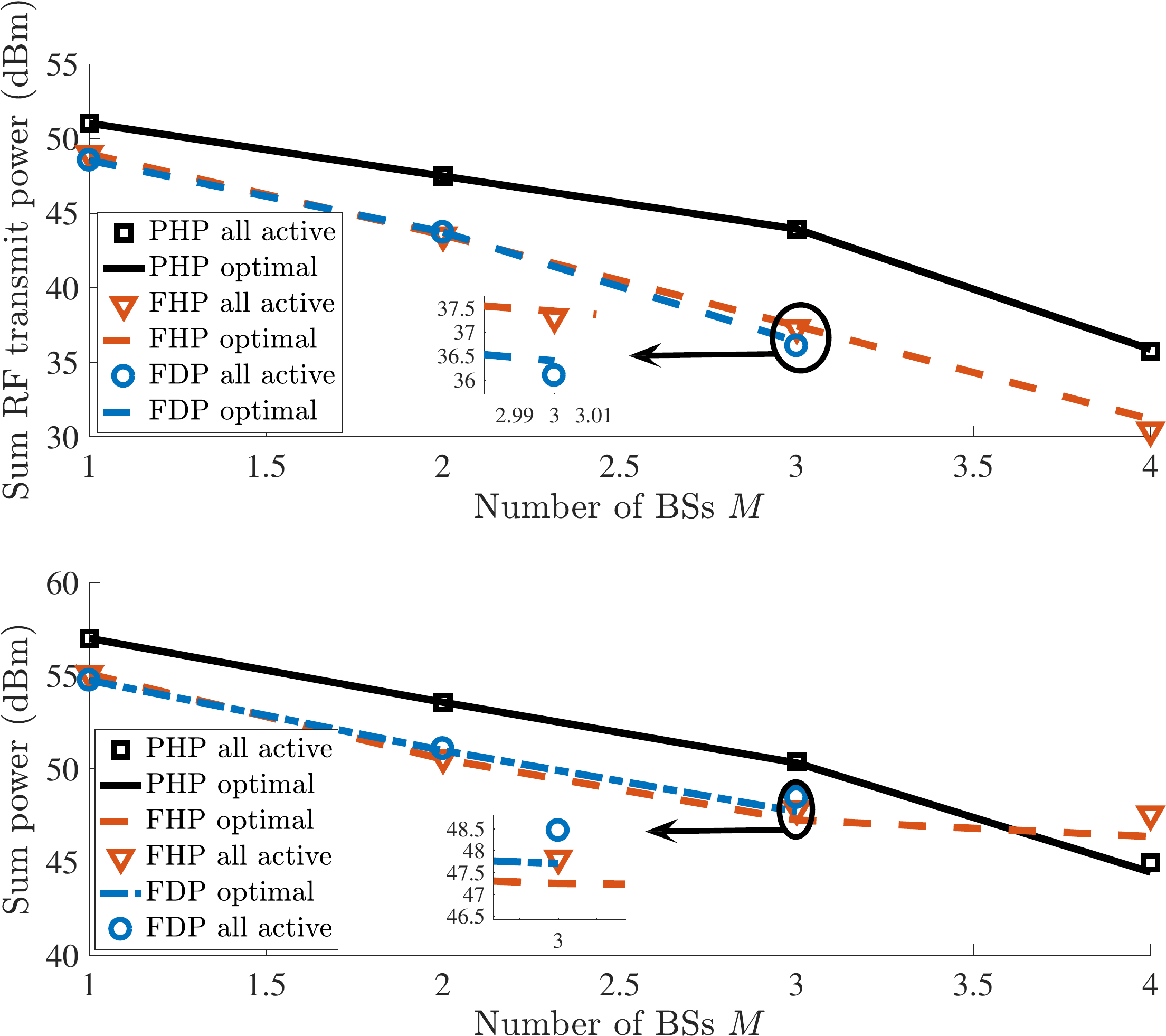}
    	\caption{(a) Sum RF transmit power and (b) sum power consumption versus the number of BSs. For each architecture, simulation results based on all BSs being active are compared to the case when the silent mode is enabled.		
    	The network parameters are the number of antennas $N_m=64$, the number of users $K=4$, the number of RF chains $L_m=4$ per BS and the target spectral efficiency $\tau_k = 4$ bit/s/Hz per user.}
    	\label{power_hl}
    \end{figure}
    
    \begin{figure}[t]
    	\centering
    	\includegraphics[width=\linewidth]{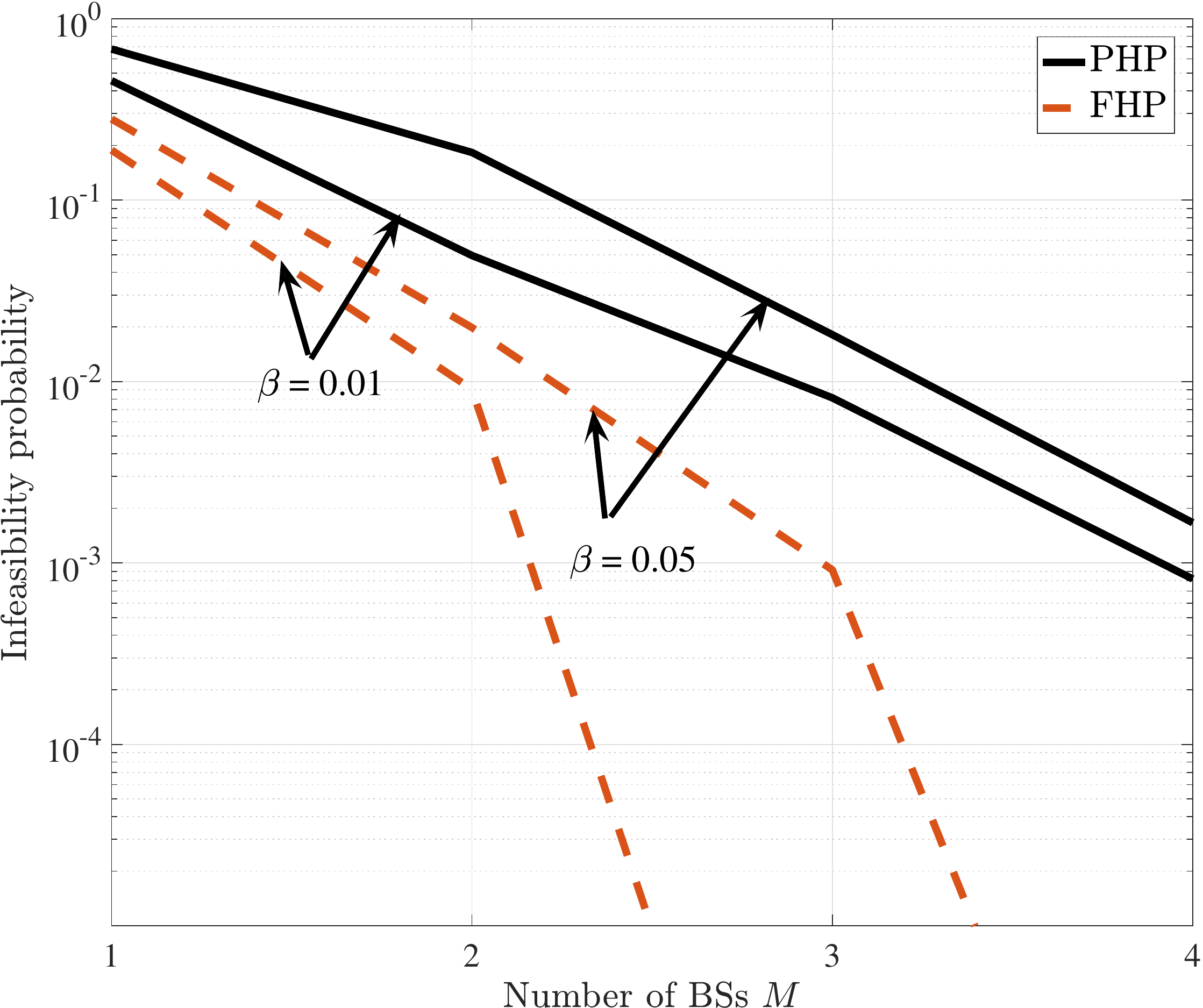}
    	\caption{Infeasibility probability versus the number of BSs. A large $\beta$ defines a denser blockage environment with less LOS transmissions.
    		The network parameters are the number of antennas $N_m=64$, the number of users $K=4$, the number of RF chains $L_m=4$ per BS and the target spectral efficiency $\tau_k = 4$ bit/s/Hz per user.}
    	\label{outage}
    \end{figure}
 
  \begin{figure}
 	\centering
 	\includegraphics[width=\linewidth]{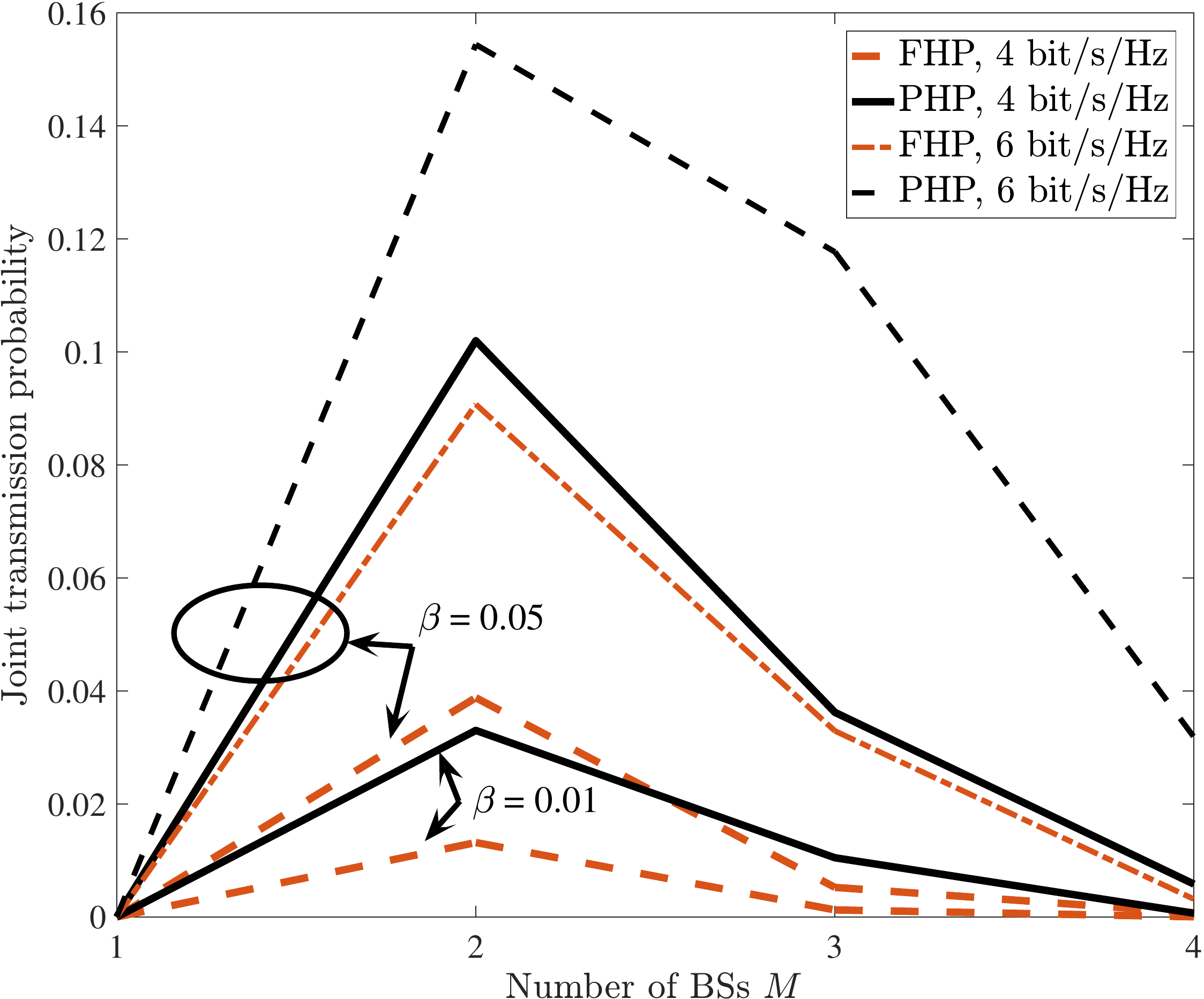}
 	\caption{BS joint transmission probability versus the number of BSs. A larger $\beta$ defines a denser blockage environment with less LOS transmissions. 4 bit/s/Hz and 6 bit/s/Hz denote the target spectral efficiency for all BSs.	
 	The network parameters are the number of antennas $N_m=64$, the number of users $K=4$ and the number of RF chains $L_m=4$ per BS.}
 	\label{coop_bs}	
 \end{figure}
 
	    Setting	the number of antennas $N_m=64$, the number of users $K=4$, the number of RF chains $L_m=4$ per BS and the target spectral efficiency $\tau_k = 4$ bit/s/Hz 
	    per user, Fig.~\ref{power_hl}(a) and \ref{power_hl}(b) show the sum RF transmit power and the sum power consumption versus the number of cooperative BSs, respectively.
	    As seen in Fig.~\ref{power_hl}(a) and \ref{power_hl}(b),
	    there is a small difference between the case with all BSs being active and the case with silent mode for all architectures. This is intuitive because when $M<4$ the average BS activation probability is close to 100\% for all architectures. For the silent mode scalar $a = 0.5$, the saved hardware power is small compared to the RF transmit power. This suggests that it suffices to consider only the case with activating all BSs with a simplified optimization problem for cooperative communications. However, in the cases with small BS activation probability and large hardware power consumption, the difference between these two cases may be more visible.

	    In Fig.~{\ref{power_hl}(a)}, the sum RF transmit power is shown to decrease with the number of BSs for all architectures. Comparing a cooperative network having 2 cooperative BSs with the cases having 1 BS, the sum RF transmit power consumption is reduced by 65\%, 71\% and 56\% for the FDP, the FHP and the PHP (see Fig.~{\ref{power_hl}(a)}), respectively.
		The result shows that the network densification and cooperative transmissions lead to a better chance for a user to be served by BSs with good channel conditions, thus, requiring less RF transmit power to achieve a target spectral efficiency\footnote{Indeed, cooperation between the BSs is at the cost of backhauling \cite{Behrooz_IAB}, which is not considered in this work.}. 
		In Fig.~\ref{power_hl}(b), the sum power consumption is reduced by 54\%, 64\% and 55\% for the FDP, the FHP and the PHP, respectively, when $M=1\rightarrow2$. For $M\ge4$, the PHP starts to consume less sum power than the FHP, which is due to the fact that the hardware increase has less effect on the sum power consumption than that of the FHP, and the sum power consumption of the PHP is dominated by the decreasing RF transmit power. 
		The results in Fig.~\ref{power_hl} indicate that, due to the network densification, cooperative transmissions have the potential to reduce both the RF transmit power and the sum power consumption of the network. Note that, in Fig.~\ref{power_hl}, we have assumed 4 users and it has reached a low RF transmit power with 4 BSs. Further increasing $M$ will increase the total power consumption as the algorithm will try to assign one BS per user for lower interference.
		For each architecture, the power consumption difference between the optimal case and the sub-optimal case increases with $M$. This is due to the fact that the BS activation probability decreases with $M$, the difference between the optimal silence strategy and sub-optimal strategy becomes more visible.

         To examine the feasibility of the proposed precoding algorithm under different blocking conditions, the infeasibility probability is illustrated in Fig.~\ref{outage} for the number of antennas $N_m=64$, the number of users $K=4$, the number of RF chains $L_m=4$ per BS and the target spectral efficiency $\tau_k = 4$ bit/s/Hz per user. Here, the infeasibility probability is defined as the probability that a feasible precoder solution cannot be found to support the target spectral efficiency. For a fixed $\beta$ which models the blockage density, the results show that the infeasibility probability drops substantially from single BS transmission to cooperative transmissions and converges to 0 as $M$ increases.
         For a given $M$, as expected, the infeasibility probability increases with $\beta$ as a result of the increased NLOS transmissions.
         The results indicate that cooperative networks are effective in reducing the infeasibility probability in areas with dense blockages and many NLOS transmissions.
        
        In Fig~\ref{coop_bs}, the joint transmission probability, defined as the probability that a user is served by more than one BS, is shown for the cases with $N_m=64$, $K=4$, $L_m=4$ per BS. For $M\ge2$, the joint transmission probability of the FHP and the PHP increases with $\beta$, as it becomes difficult for a single BS to satisfy the target spectral efficiency when the probability of NLOS transmission is high.
        For $\beta = 0.05$, as expected, the joint transmission probability of FHP and PHP increases with the target spectral efficiency. Thus, for high probability of NLOS transmission and high spectral efficiency the users need more joint transmissions. 
        Also, we notice that the joint transmission probability decreases when the number of BSs approaches the number of users. Intuitively, this is because joint transmissions from multiple BSs impose additional interference to other users and the algorithm tries to assign one BS per user. Even though the joint transmission probability is small, BS joint transmission still provides performance gain in the following two aspects. 
        First, by jointly performing the user association, a user is optimally associated with BSs having good channel conditions and least transmit power that satisfies the quality-of-service constraints. Second, the BS silence strategy reduces the average activation probability for larger number of cooperating BSs, thus optimizing the sum power consumption. 
        Additionally, Fig.~\ref{coop_bs} shows that, for the same $\beta$, the joint transmission probability of the PHP is higher than that of the FHP, since the infeasibility probability of the PHP is larger, thus requiring more help from other BSs.

        \begin{figure}[t]
        	\centering
        	\includegraphics[width=\linewidth]{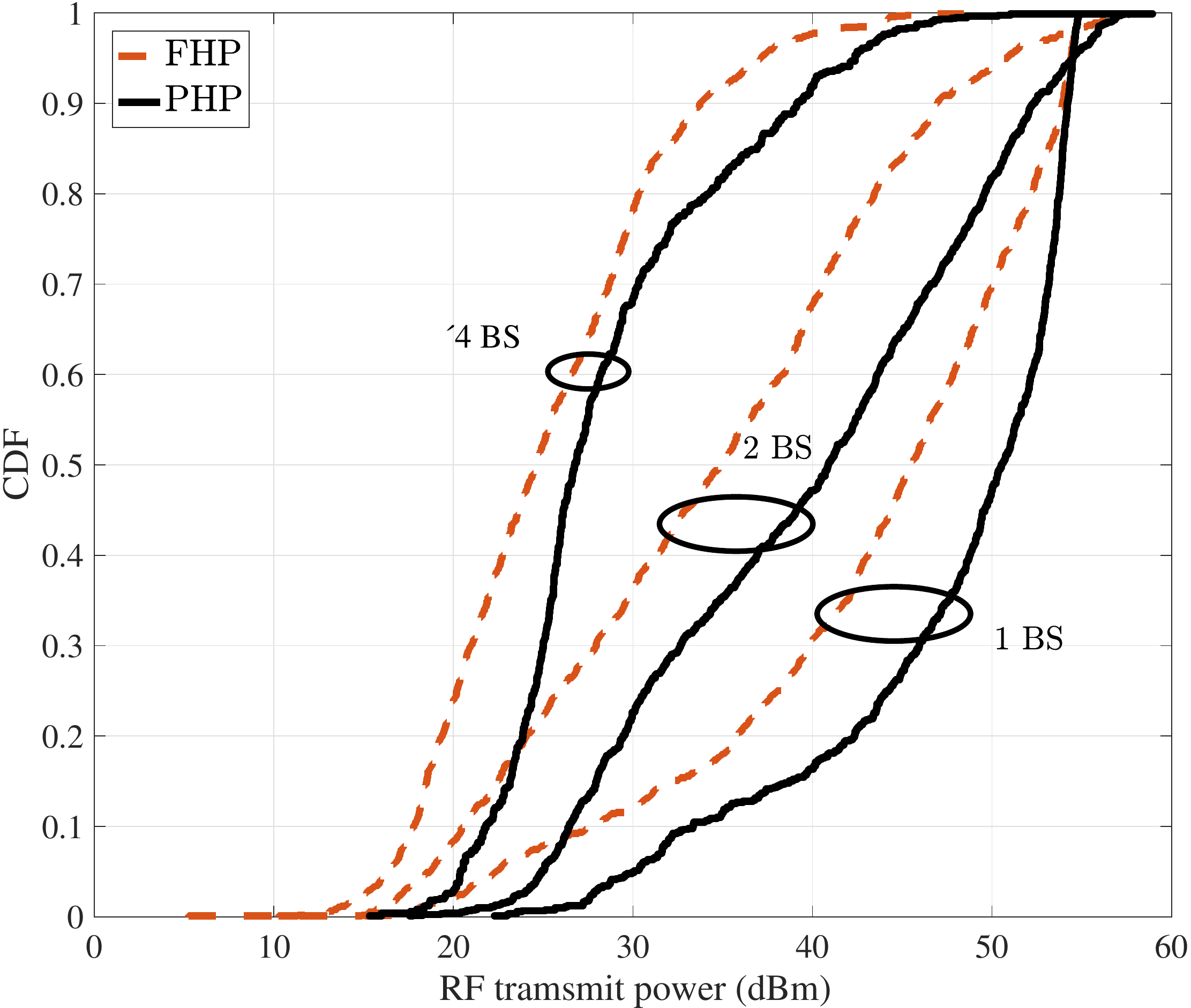}
        	\caption{CDF of the average RF transmit power for individual BSs. The network parameters are set to $N_m=64$, $K=4$, $L_m=4$ per BS and $\tau_k = 4$ bit/s/Hz per user.}
        	\label{powerCDF_bs}	
        \end{figure}	
        
	    To check the effect of number of BSs on the RF transmit power range, Fig.~\ref{powerCDF_bs} shows the RF transmit power cumulative distribution function of individual BSs for $M= \{1, 2, 4\}, N_m=64, K=L_m=4, \forall m,$ and $\tau_k = 4$ bit/s/Hz, $\forall k$. In general, cooperative transmissions reduce the transmit power variations by optimizing the user associations, which might be very useful in Effective Isotropic Radiated Power (EIRP) limited deployment scenarios.
        In Fig.~\ref{powerCDF_bs}, the RF transmit power for the case with 4 BSs has less variance than the cases with smaller number of BSs. The 95-th percentile of the transmit power varies from 37 dBm in the 4-BS cooperation case to 54 dBm in the 1-BS case for FHP. 
	    The power variation may affect the performance of hardware components  such as power amplifiers considering the output power dynamic range.
	    
	    The simulations in this subsection verify that a cooperative network with more BSs can achieve lower RF/sum power consumption, and is effective in reducing the infeasibility probability and transmit power variations. However, we also notice that for channels with high mobility users/environments, BS cooperation increases the difficulty of backhauling and coordination overhead \cite{Behrooz_IAB}. Studying such scenarios and  their effects on the network performance is however out of the scope of this paper, and is left for future work.
        \subsection{On the OFDM systems}
        \label{ofdm_sim}
  \begin{figure}[t]
	\centering
	\includegraphics[width=\linewidth]{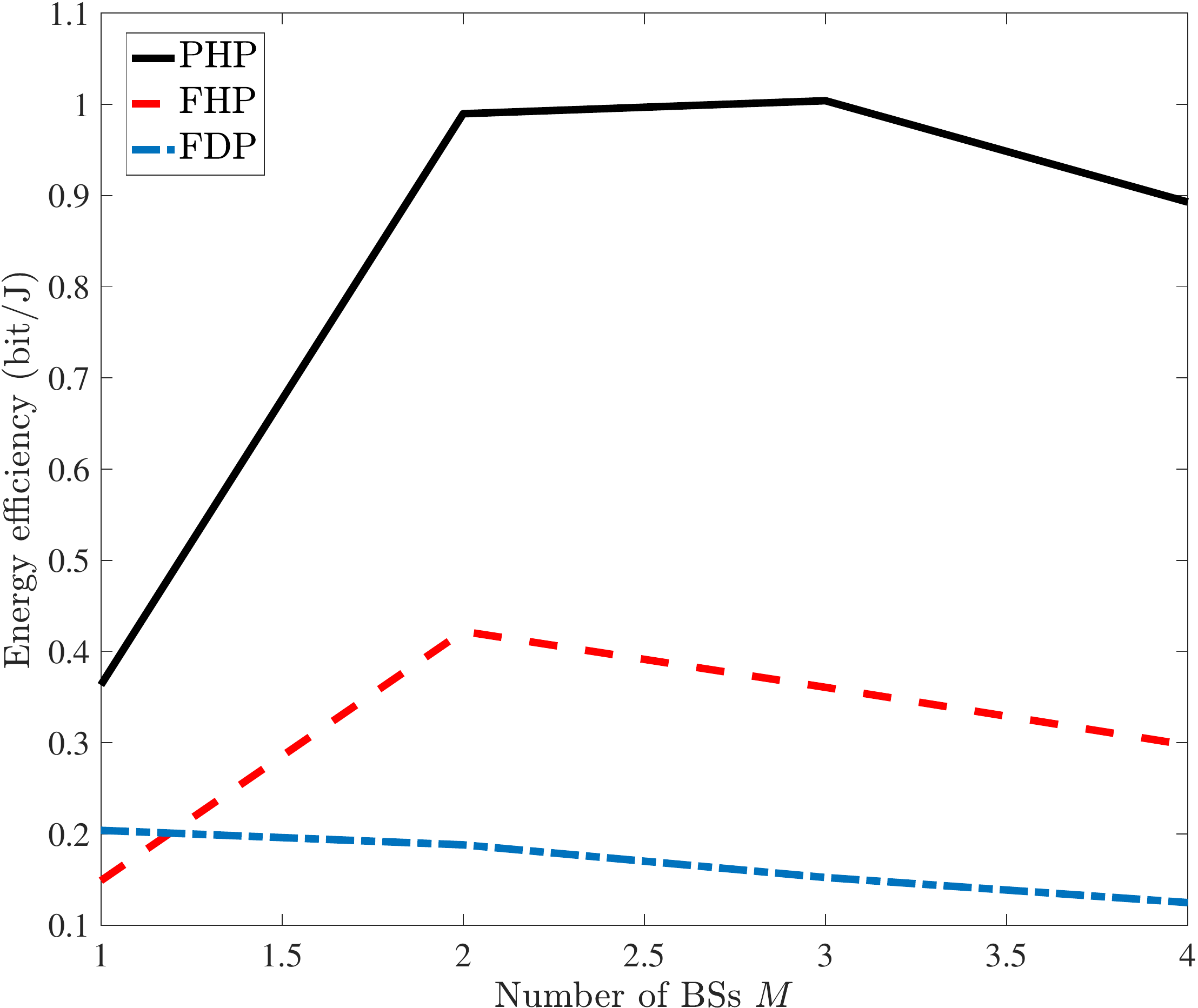}
	\caption{Energy efficiency versus the number of BSs. For each architecture, simulation results are based on all BSs being active.		
		The network parameters are the number of antennas $N_m=64$, the number of users $K=4$ and the number of RF chains $L_m=4$ per BS.}
	\label{EE_OFDM}
\end{figure}

In this subsection, we present simulation results of cooperative hybrid precoding for the OFDM system case. We compare the energy efficiency among PHP, FDP and FHP versus per sub-carrier target spectral efficiency and the number of BSs. We use the following metric to quantify the energy efficiency of the OFDM based cooperative system: 
\ieb
&& E = \frac{\sum_{k=1}^{K} \sum_{n_{\mathrm{s}}=0}^{N_{\mathrm{s}}-1} \Gamma_k[n_s]}{\sum_{m=1}^{M}P_m},
\ien
which is the ratio of the sum of the achievable spectral efficiency to the sum of BS power consumption. 

In Fig.~\ref{EE_OFDM}, the value of cooperation is shown for OFDM systems with $N_{\mathrm{s}}= 64, K=4, P_{\mathrm{PS}}= 40$ mW, $N_m=64, L_m=4, \forall m$, the per sub-carrier target spectral efficiency set to 4 bit/s/Hz, and with the bandwidth per sub-carrier set to 3 MHz. In Fig.~\ref{EE_OFDM}, the energy efficiency is shown for $M={1,2,3,4}$. For FHP and PHP with $M\le2$, the energy efficiency improves due to the reduced average power consumption and the less number of unused sub-carriers. However, as more BSs join the network, the hardware power consumption increases with $M$ and the spectrum efficiency has no significant increase, further increasing $M$ reduces the energy efficiency.

\vspace{-2mm}
\section{Conclusion}
In this paper, we proposed hybrid beamforming algorithms that enable joint transmissions in a cooperative multi-cell multi-user mmWave network, for both FHP and PHP, for single-carrier and OFDM systems.
The proposed algorithm allows to minimize the total RF and hardware power consumption under per-user target spectral efficiency constraint and per-BS maximum RF transmit power constraint, and finds the optimal user associations and the BS silence strategy. 
By allowing joint transmissions from
multiple BSs to each user, we first showed that the joint analog and digital precoding problems can be decoupled
into independent equal-gain transmission problems and relaxed convex semidefinite programs.
Then, we analyzed the Lagrangian dual problem of the convex digital precoder optimization problem which gave the conditions for the optimal user association strategy that minimizes the sum power consumption of the network.
Next, based on the convex envelope of the objective function of the hybrid beamforming algorithm, we proposed a sub-optimal hybrid precoding algorithm in terms of the silence strategy with low complexity.
 Simulations on the power consumption verifies that the FHP achieves similar RF transmit power compared to the FDP and the hardware power consumption has a large impact on the sum power consumption of the hybrid precoding architectures.
Furthermore, cooperative transmissions are shown to increase the energy efficiency and reduce the sum power consumption of the network, the infeasibility probability, and the RF transmit power variations.

	 \bibliographystyle{IEEEtran} %lic.bst is the style file
	\bibliography{JointPrecoding}
	\vspace{5cm}
		\begin{IEEEbiography}[{\includegraphics[width=1in,clip,keepaspectratio]{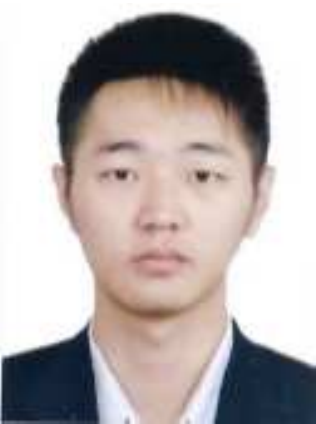}}]{Chao Fang} received the M.S. degree in electrical engineering from the Chalmers University of
			Technology, Gothenburg, Sweden, in 2015, where
			he is currently pursuing the Ph.D. degree with the
			Communication Systems Group, Department of Electrical Engineering. His research interests include heterogeneous cellular networks, integrated access and backhaul, and millimeter-wave
			communications.
	\end{IEEEbiography}
	\vskip 0pt plus -1fil
	\begin{IEEEbiography}[{\includegraphics[width=1in,clip,keepaspectratio]{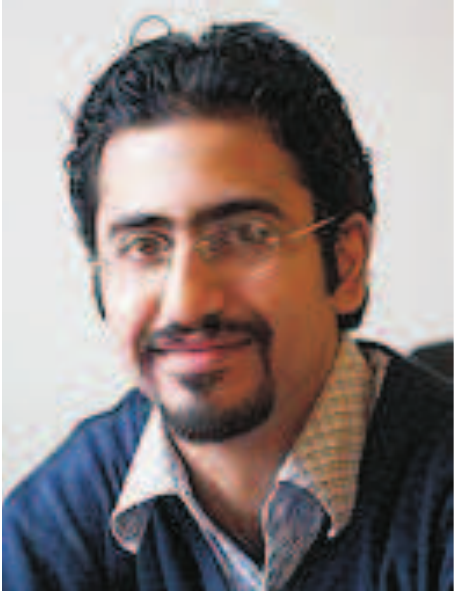}}]{Behrooz Makki} [M'19, SM'19] received his
		PhD degree in Communication Engineering from Chalmers University of Technology, Gothenburg, Sweden. In 2013-2017, he was a Postdoc researcher
		at Chalmers University. Currently, he works as Senior Researcher in Ericsson Research, Gothenburg, Sweden.
		
		Behrooz is the recipient of the VR Research Link grant, Sweden, 2014, the Ericsson’s Research grant, Sweden, 2013, 2014 and 2015, the ICT SEED grant,
		Sweden, 2017, as well as the Wallenbergs research grant, Sweden, 2018. Also, Behrooz is the recipient of the IEEE best reviewer
		award, IEEE Transactions on Wireless Communications, 2018. Currently, he works as an Editor in IEEE Wireless Communications Letters, IEEE
		Communications Letters, the journal of Communications and Information Networks, as well as the Associate Editor in Frontiers in Communications and
		Networks. He was a member of European Commission projects “mm-Wave based Mobile Radio Access Network for 5G Integrated Communications”
		and “ARTIST4G” as well as various national and international research collaborations. His current research interests include integrated access and
		backhaul, Green communications, millimeter wave communications, finite block-length analysis and backhauling. He has co-authored 64 journal papers, 46 conference papers and 60 patent applications.
	\end{IEEEbiography}
	\vskip 0pt plus -1fil
	\begin{IEEEbiography}
	[{\includegraphics[width=1in,clip,keepaspectratio]{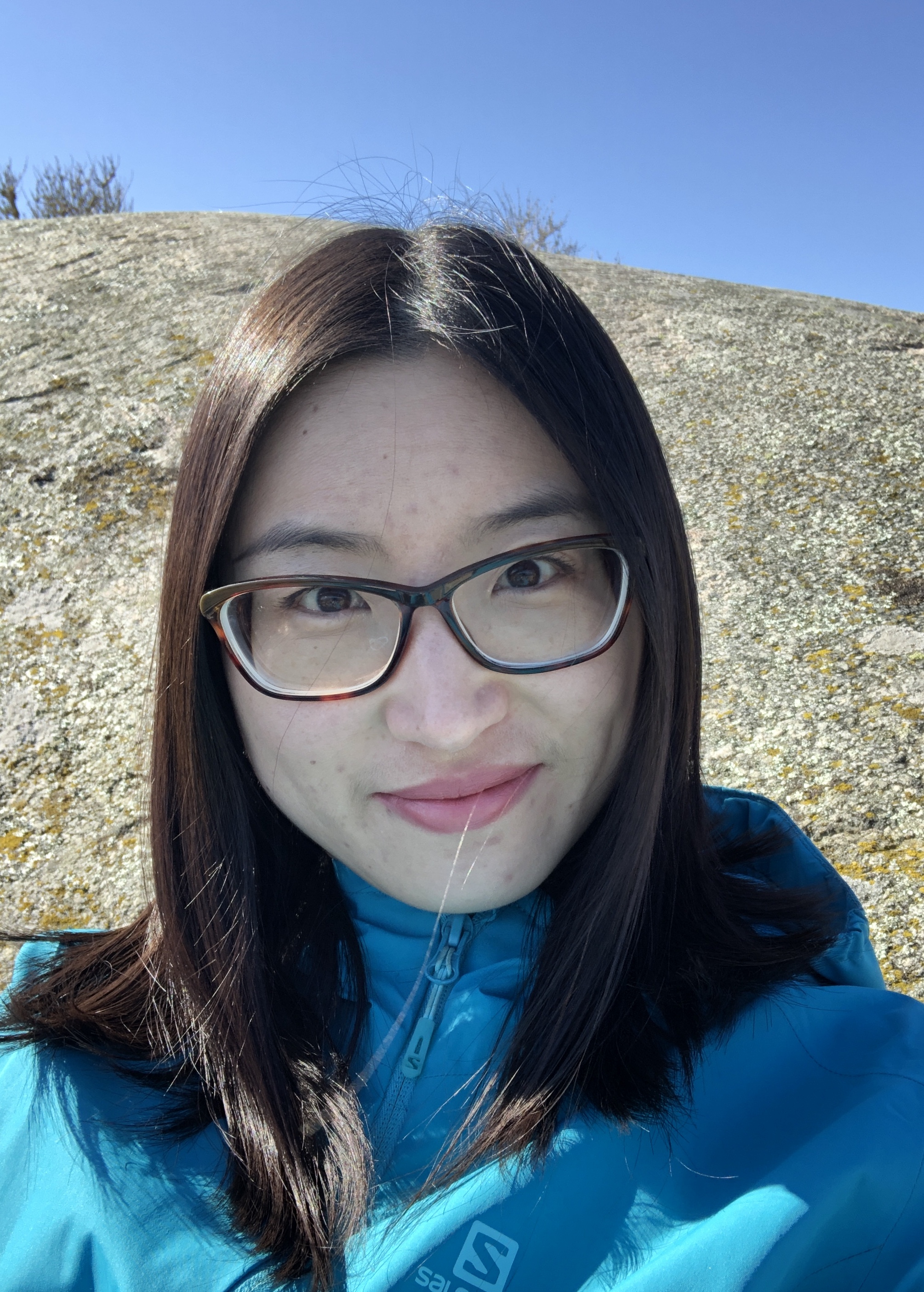}}]{Jingya Li} is a senior researcher at Ericsson Research working with 5G and future radio access technologies. She leads 5G research and standardization in the areas of public safety mission critical communications. She is a key contributor to 5G NR initial access, RIM and CLI, V2X and LTE latency reduction standards specifications within 3GPP RAN WG1. She was the recipient of the IEEE 2015 ICC Best Paper Award and IEEE 2017 Sweden VT-COM-IT Joint Chapter Best Student Journal Paper Award. She holds a Ph.D. degree (2015) in Electrical Engineering from Chalmers University of Technology, Gothenburg, Sweden.
	\end{IEEEbiography}    
	\vskip 0pt plus -1fil
	\begin{IEEEbiography}
		[{\includegraphics[width=1in,clip,keepaspectratio]{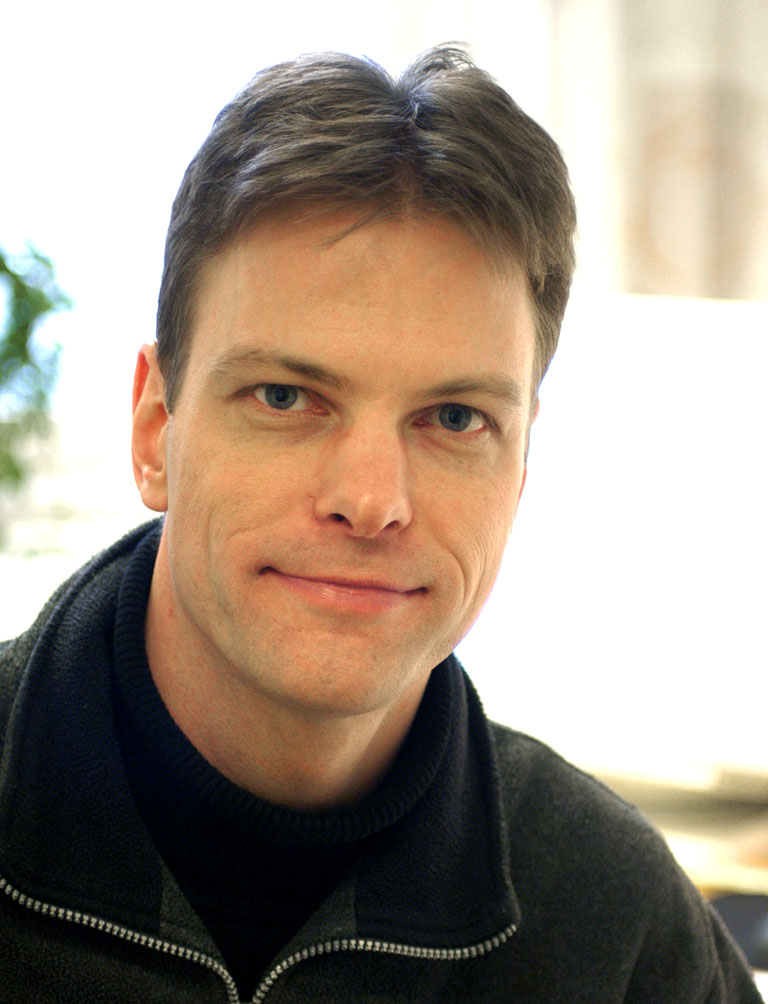}}]{ Tommy Svensson} [S’98, M’03, SM’10] is full Professor in Communication Systems at Chalmers University of Technology in Gothenburg, Sweden, where he is leading the Wireless Systems research on air interface and wireless backhaul networking technologies for future wireless systems. He received a Ph.D. in Information theory from Chalmers in 2003, and he has worked at Ericsson AB with core networks, radio access networks, and microwave transmission products. He was involved in the European WINNER and ARTIST4G projects that made important contributions to the 3GPP LTE standards, the EU FP7 METIS and the EU H2020 5GPPP mmMAGIC and 5GCar projects towards 5G and currently the Hexa-X, RISE-6G and SEMANTIC projects towards 6G, as well as in the ChaseOn antenna systems excellence center at Chalmers targeting mm-wave and (sub)-THz solutions for 5G/6G access, backhaul/ fronthaul and V2X scenarios. His research interests include design and analysis of physical layer algorithms, multiple access, resource allocation, cooperative systems, moving networks, and satellite networks. He has co-authored 5 books, 94 journal papers, 129 conference papers and 54 public EU projects deliverables. He is Chairman of the IEEE Sweden joint Vehicular Technology/ Communications/ Information Theory Societies chapter, founding editorial board member and editor of IEEE JSAC Series on Machine Learning in Communications and Networks, has been editor of IEEE Transactions on Wireless Communications, IEEE Wireless Communications Letters, Guest editor of several top journals, organized several tutorials and workshops at top IEEE conferences, and served as coordinator of the Communication Engineering Master’s Program at Chalmers.
	\end{IEEEbiography}  
\end{document}